\documentclass[a4paper,11pt]{article}
\pdfoutput=1
\usepackage{jheppub}
\usepackage{amsmath, amsfonts, amssymb}
\usepackage{graphicx, subfig, hyperref, feynmp-auto}

\newcommand{\arrowhack}[5]{%
  \fmfcmd{%
  vardef middir(expr p,ang) =
   dir(angle direction length(p)/2 of p + ang)
  enddef;
  style_def marrow#1 expr p =
   shrink(.7);
   cfill(arrow p shifted(#2thick*middir(p,90)));
   endshrink
  enddef;}
  \fmf{marrow#1, label=#3, label.dist=#4}{#5}}

\title{Bootstrapping the long-range Ising model in three dimensions}
\preprint{YITP-SB-18-27}
\author[a]{Connor Behan}
\affiliation[a]{C. N. Yang Institute for Theoretical Physics, Stony Brook University, \\ Stony Brook, NY 11794, USA}
\emailAdd{connor.behan@gmail.com}

\abstract{The 3D Ising model and the generalized free scalar of dimension at least 0.75 belong to a continuous line of nonlocal fixed points, each referred to as a long-range Ising model. They can be distinguished by the dimension of the lightest spin-2 operator, which interpolates between 3 and 3.5 if we focus on the non-trivial part of the fixed line. A property common to all such theories is the presence of three relevant conformal primaries, two of which form a shadow pair. This pair is analogous to a superconformal multiplet in that it enforces relations between certain conformal blocks. By demanding that crossing symmetry and unitarity hold for a set of correlators involving the relevant operators, we compute numerical bounds on their scaling dimensions and OPE coefficients. Specifically, we raise the minimal spin-2 operator dimension to find successively smaller regions which eventually form a kink. Whenever a kink appears, its co-ordinates show good agreement with the epsilon expansion predictions for the critical exponents in the corresponding statistical model. As a byproduct, our results reveal an infinite tower of protected operators with odd spin.}
\keywords{conformal field theory, conformal bootstrap}

\begin{document}
\maketitle
\flushbottom

\section{Introduction}
\label{sec:intro}
In recent years, many strongly interacting quantum field theories, once considered intractable, have begun to yield to a technique known as the conformal bootstrap. This is the old idea of using crossing symmetry of the four-point function to constrain the operator algebra of a conformal field theory (CFT) \cite{fgg73,p74}. Brought on by a renewed understanding of conformal blocks \cite{do00, do03, do11}, a modern numerical bootstrap, capable of bounding operator dimensions and three-point function coefficients was formulated in \cite{rrtv08}. Some notable reviews are \cite{r16a, s16a, prv18}.

The 2D Ising model, solved exactly in \cite{o44}, was the first theory observed to saturate these bounds. It was later seen that this also applies to the 3D Ising model and subsequent work led to the most precise determination of its critical exponents \cite{epprsv12, epprsv14, kps14, kpsv16}. On the other hand, long-range Ising (LRI) models, defined by relaxing the requirement for spins to interact with nearest neighbours, have not been solved in any dimension, despite exhibiting critical behaviour in all $1 \leq d < 4$ \cite{d69}. These are prime candidates for a non-perturbative bootstrap study. The models we consider will have ferromagnetic ($J > 0$) interactions that fall off as a power-law:
\begin{equation}
H_{\mathrm{LRI}} = -J \sum_{i, j} \frac{\sigma_i \sigma_j}{|i - j|^{d + s}} \; . \label{lri-hamiltonian}
\end{equation}
On the fixed line parameterized by $s$, two regimes are well understood. If $s$ exceeds a certain crossover value, which we call $s_*$, the critical exponents become those of the (short-range) Ising model. For the critical exponents to become those of mean-field theory instead, we must have $s < \frac{d}{2}$. The most interesting range is therefore $\frac{d}{2} < s < s_*$. Choosing several such values of $s$ and $d = 3$, our goal is to do for these models what \cite{epprsv12, epprsv14} did for their short-range counterpart.

Monte Carlo simulations, which become more expensive as $d$ is increased, have yet to produce results for the 3D long-range Ising model. Even the 2D LRI has only been simulated relatively recently \cite{lb02, p12, bpr13, apr14, hst17}. For this reason, our approach will be heavily guided by perturbation theory. The perturbation theory around $s = \frac{d}{2}$ goes back to \cite{fmn72}, with significant improvements in \cite{s73}. A perturbative approach to $s = s_*$ was recently developed in \cite{brrz17a, brrz17b} and could benefit from further non-perturbative checks. The theory in between these endpoints should be conformal, by the arguments in \cite{prvz16}, and lack a stress-energy tensor. This does not conflict with the simplest motivation for conformal covariance which is based on Ward identities. The presence of a tracless stress-energy tensor is only a \textit{sufficient} condition for having a CFT.\footnote{Without a stress-energy tensor, we can have a theory satisfying all of the usual CFT axioms except locality. Although we will continue to call this object a CFT, it is also commonly referred to as a conformal theory or CT \cite{s16b, pptvv16}.}

While nonlocal bootstrap constraints are interesting to explore in their own right, an intriguing possibility is that the LRI could be realized in observable systems \cite{r05}. The most satisfying argument would derive an effective $V(r) \sim \frac{1}{r^{d + s}}$ potential in a naturally occuring crystal. Doing this from first principles, \textit{e.g.} using the even longer ranged $V(r) \sim \frac{1}{r}$ that holds microscopically, would be a triumph of many-body physics. Rather, we will content ourselves with taking a look at the toy-model involving two hydrogenic atoms.

In the setup for this classic problem \cite{p28}, we have protons at positions $\textbf{r}_A$, $\textbf{r}_B$ and electrons at positions $\textbf{r}_1$, $\textbf{r}_2$. The Hamiltonian $H = H_0 + H_{\mathrm{int}}$ is given by
\begin{eqnarray}
H_0 &=& \frac{p_1^2}{2m} + \frac{p_2^2}{2m} - \frac{e^2}{r_{1A}} - \frac{e^2}{r_{2B}} \nonumber \\
H_{\mathrm{int}} &=& e^2 \left ( \frac{1}{r_{AB}} + \frac{1}{r_{12}} - \frac{1}{r_{1B}} - \frac{1}{r_{2A}} \right ) \; . \label{h2-hamiltonian}
\end{eqnarray}
Corrections to the energy can be found perturbatively from the symmetric and anti-symmetric ground-states of $H_0$.
\begin{equation}
\psi_{\pm}(\textbf{r}_1, \textbf{r}_2) = \frac{1}{\sqrt{2 \pm 2 \zeta^2}} \left [ \psi_A(\textbf{r}_1)\psi_B(\textbf{r}_2) \pm \psi_A(\textbf{r}_2)\psi_B(\textbf{r}_1) \right ] \;\; , \;\; \psi_I(\textbf{r}_i) \equiv \frac{1}{\sqrt{\pi a^3}} e^{-\frac{r_{iI}}{a}} \label{h2-wavefunctions}
\end{equation}
Expressing the inter-atomic distance in units of the Bohr radius, one can check that
\begin{equation}
\zeta = e^{-\rho} \left ( 1 + \rho + \frac{1}{3} \rho^2 \right ) \;\; , \;\; \rho \equiv \frac{r_{AB}}{a} \label{h2-normalization}
\end{equation}
is the factor that appears in the normalization. Further integration reveals the following first-order shift.
\begin{eqnarray}
E^{(1)}_\pm &=& \left < \psi_\pm \right | H_{\mathrm{int}} \left | \psi_\pm \right > = \frac{e^2}{r_{AB}} + \frac{e^2}{a} \frac{I_1 \pm I_2}{1 \pm \zeta^2} \label{shift-sr} \\
I_1 &=& -\frac{1}{\rho} + e^{-2\rho} \left ( \frac{1}{\rho} + \frac{5}{8} - \frac{3}{4} \rho - \frac{1}{6} \rho^2 \right ) \nonumber \\
I_2 &=& -2 \zeta e^{-\rho} (1 + \rho) + e^{-2\rho} \left ( \frac{5}{8} - \frac{23}{20} \rho - \frac{3}{5} \rho^2 - \frac{1}{15} \rho^3 \right ) \nonumber \\
&& + \frac{6}{5\rho} \left [ \zeta^2 (\gamma + \log \rho) - 2 \left ( 1 - \frac{1}{3} \rho^2 + \frac{1}{9} \rho^4 \right ) \mathrm{Ei}(-2\rho) + e^{2\rho} \left ( 1 - \rho + \frac{1}{3} \rho^2 \right )^2 \mathrm{Ei}(-4\rho) \right ] \nonumber
\end{eqnarray}
The sign of $I_2$ tells us that $E_+^{(1)} < E_-^{(1)}$. As the spatial wavefunction $\psi_+$ must be paired with an anti-symmetric spin part, this contribution to the potential is anti-ferromagnetic. Another interesting property of \eqref{shift-sr} is its exponential decay for large $\rho$. This result, which happens to be a 3D coincidence \cite{is15}, is much stronger than what follows from dimensional analysis.

For a less symmetric state, the $\alpha^n$ term would decay as $\rho^{-3n}$ according to the multipole expansion
\begin{equation}
H_{\mathrm{int}} = \frac{\textbf{r}_{1A} \cdot \textbf{r}_{2B} - 3(\textbf{r}_{1A} \cdot \hat{\textbf{r}}_{AB})(\textbf{r}_{2B} \cdot \hat{\textbf{r}}_{AB})}{r_{AB}^3} + O \left ( \frac{1}{r_{AB}^4} \right ) \; , \label{multipole}
\end{equation}
and the first-order $E^{(1)}$ would always be dominant. The solution \eqref{shift-sr}, on the other hand, is special because the only surviving terms are non-perturbative in $r_{AB}^{-1}$. It is therefore necessary to compute $E^{(2)}$ to find the leading contribution at large distances. This potential, called the van der Waals interaction, is proportional to $\alpha^2 / r_{AB}^6$. If this force between two atoms were to appear again in an infinite lattice, it would justify a Hamiltonian like \eqref{lri-hamiltonian} with $s = 3$. This is still not enough to get out of the short-range universality class. As we will explain shortly, the crossover value is given by
\begin{equation}
s_* = d - 2\Delta_\sigma^{\mathrm{SRI}} \; , \label{s-star}
\end{equation}
which is about $1.96$ in three dimensions. Nevertheless, different values of $s$ above $s_*$ can be distinguished from finite-size effects \cite{dr01}.

There are indeed systems that are known to have significant next-nearest neighbour interactions even though the precise power has yet to be measured. These include some high-temperature magnetic and spin-ice materials \cite{wtchfpgsl16, bhdggmwccmf01}, in which an anti-ferromagnetic short-range force is overcome by a ferromagnetic long-range force, as well as magnetic thin films where the competition is reversed \cite{dmw00}. Tunable long-range forces, between both Ising and multi-component spins, have recently been achieved in quantum simulators based on trapped ions \cite{bskwfubb12, bsbwrfb16} and cold atoms \cite{lhdlmde16, hgck16}. While these are useful for studying quantum phase transitions close to zero temperature, it is likely that driving the system to a thermal phase transition would introduce a radical departure from a Hamiltonian like \eqref{lri-hamiltonian}.\footnote{A bootstrap approach with possible applications to this scenario was recently developed in \cite{ikmps18, gmnw18, ps18}.} Our bootstrap of the LRI is therefore somewhat exotic despite its status as a natural starting point for understanding the space of nonlocal theories.

This paper is organized as follows. In section \ref{sec:continuum}, we discuss what is known about the fixed line including some non-renormalization theorems that are crucial for the bootstrap. In the process we review, and slightly extend, the perturbative literature. In section \ref{sec:shadow}, we discuss two ways in which the LRI bootstrap is structurally similar to that of a superconformal theory. In particular, we show that pairs of global conformal blocks can be packaged together and present a new non-renormalization theorem that affects infinitely many operators. In section \ref{sec:numerics}, we present new numerical constraints, starting with the ones accessible to a single correlator and working our way up to six. The six-correlator bounds show features in places where we expect long-range Ising models to live. We conclude in section \ref{sec:conclusion} and present explicit crossing equations in Appendix \ref{sec:appa}.

\section{The continuum approach}
\label{sec:continuum}
To study \eqref{lri-hamiltonian} with the renormalization group, \cite{fmn72} introduced the action
\begin{equation}
S = \int -\frac{1}{2} \phi \partial^s \phi + \frac{\lambda}{4!} \phi^4 \textup{d}x \label{flow1-action}
\end{equation}
with a $\mathbb{Z}_2$ global symmetry.\footnote{The fractional derivative is a shorthand for the nonlocal operator that acts as $\partial^s \phi(x) \equiv \int \frac{\phi(y)}{|x - y|^{d + s}} \textup{d}y$.} The kinetic term, with momentum space propagator $|k|^{-s}$, is exactly the action for a generalized free scalar of dimension $\frac{d - s}{2}$. The $\phi^4$ perturbation drives the system to the LRI fixed point in the spirit of Wilson-Fisher. When $s > \frac{d}{2}$, this interaction is relevant which makes the fixed point non-trivial. It is therefore natural to compute observables as an expansion in $\varepsilon \equiv 2s - d$. In terms of this parameter, the beta function is
\begin{equation}
\beta(\lambda) = -\varepsilon \lambda + \frac{3\lambda^2}{\Gamma \left ( \frac{d}{2} \right ) (4\pi)^{\frac{d}{2}}} + O(\lambda^3) \; . \label{flow1-beta}
\end{equation}

\subsection{Two non-renormalized dimensions}
\begin{figure}[h]
\centering
\begin{fmffile}{graph1}
\begin{fmfgraph*}(100, 60)
\fmfleft{p1}
\fmfright{p2}
\fmf{plain}{p1,i1,i2,p2}
\fmf{plain,left}{i1,i2,i1}
\fmflabel{k}{p1}
\fmflabel{k}{p2}
\end{fmfgraph*}
\end{fmffile}
\vspace{-1cm}
\caption{The first non-vanishing correction to the two-point function of $\phi$.}
\label{graph-phi}
\end{figure}
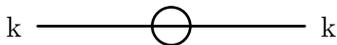
An essential property of the model is the observation that the propagator does not receive loop corrections. The intuitive justification for this is that a nonlocal kinetic term cannot be modified by local divergences. A useful test is evaluating the first two-loop diagram that could potentially influence $\gamma_\phi$. Choosing an analytic regularization scheme, the diagram in Figure \ref{graph-phi} yields
\begin{eqnarray}
G(k) &=& \frac{\lambda^2}{6(4\pi)^d} \frac{\Gamma \left ( \frac{3s}{2} - d \right ) \Gamma \left ( \frac{d - s}{2} \right )^3}{\Gamma \left ( \frac{3d - 3s}{2} \right ) \Gamma \left ( \frac{s}{2} \right )^3} \left | \frac{k}{\mu} \right |^{2d - 3s} \nonumber \\
&=& \frac{\lambda^2}{6(4\pi)^d} \frac{\Gamma \left ( \frac{3\varepsilon - d}{4} \right ) \Gamma \left ( \frac{d - \varepsilon}{4} \right )^3}{\Gamma \left ( \frac{3d - 3\varepsilon}{4} \right ) \Gamma \left ( \frac{d + \varepsilon}{4} \right )^3} \left | \frac{k}{\mu} \right |^{\frac{d - 3\varepsilon}{2}} \; . \label{two-loop-phi}
\end{eqnarray}
This requires no minimal subtraction as the $\varepsilon \rightarrow 0$ limit is finite for all $1 \leq d < 4$. Indeed, $\Gamma \left ( -\frac{d}{4} \right )$ is the only gamma function with a non-positive argument. It is reassuring that a pole appears for $d = 4$, which is precisely the value that turns \eqref{flow1-action} into a local theory.\footnote{Because \eqref{two-loop-phi} is a two-loop contribution, we must be careful when using it to solve for $\gamma_\phi$ in the Wilson-Fisher fixed point. Setting $d = 4$ in the second line is not valid. We must instead set $(s, d) = (2, 4 - \varepsilon)$ in the first line. This diagram therefore provides a counter-example to the widely discussed ``effective dimension'' idea \cite{lkmy10, bfmy12, bpr13, apr14}, which we do not find convincing.} The upshot is that the dimension of $\phi$ has the exact expression
\begin{equation}
\Delta_\phi = \frac{d - s}{2} \; , \label{non-renorm1}
\end{equation}
a relation which has been proven rigorously in \cite{lsw17}. By demanding that the short-range scaling dimensions are approached continuously, we can take $\Delta_\sigma^{\mathrm{SRI}} = \frac{d - s_*}{2}$ as the definition of $s_*$. This is a correction to \cite{fmn72} which initially predicted $s_* = 2$. The correct behaviour at the crossover was explicitly demonstrated in \cite{s73} by taking $d \rightarrow 4$ to make the entire flow perturbative. When this is done, one sees that the weakly irrelevant operator $\phi \partial^2 \phi$ is responsible for the breakdown of \eqref{non-renorm1}.

The second operator with an exactly known scaling dimension is $\phi^3$. This is obvious in the Wilson-Fisher fixed point as the equation of motion $\partial^2 \phi = \frac{\lambda}{3!} \phi^3$ places $\phi^3$ squarely in the $\phi$ multiplet.\footnote{The implications of this for anomalous dimensions were explored in \cite{rt15}.} In the LRI, this protected dimension is instead a consequence of the \textit{nonlocal} equation of motion
\begin{equation}
\partial^s \phi = \frac{\lambda}{3!} \phi^3 \; , \label{nonlocal-eom}
\end{equation}
with $\phi^3$ remaining an independent primary. From \eqref{nonlocal-eom}, we read off
\begin{equation}
\Delta_{\phi^3} = \frac{d + s}{2} \; . \label{non-renorm2}
\end{equation}
These operators, satisfying $\Delta_\phi + \Delta_{\phi^3} = d$, are often said to form a \textit{shadow pair} \cite{prvz16}.

\subsection{More anomalous dimensions}
\begin{figure}[h]
\centering
\begin{fmffile}{graph2}
\begin{fmfgraph*}(160, 100)
\fmfleft{v1}
\fmfright{v2}
\fmftop{t}
\fmf{plain}{v1,v,v2}
\fmf{plain}{v,v}
\fmfforce{0.5w,1.04h}{t}
\fmfdot{t}
\fmffreeze

\fmftop{a1}
\fmfbottom{a2}
\fmfforce{0.5w,0.5h}{a2}

\arrowhack{a}{-4}{$k_1$}{10}{v1,v}
\arrowhack{b}{-4}{$k_1 + k_2$}{10}{v,v2}
\arrowhack{c}{12}{$p$}{-30}{a2,a1}
\arrowhack{d}{12}{$k_2 + p$}{-55}{a1,a2}
\end{fmfgraph*}
\end{fmffile}
\vspace{-1cm}
\caption{The $T_{\mu\nu}$ operator, represented by a dot at the top with momentum flowing in, has its legs saturated by $\phi$.}
\label{graph-t1}
\end{figure}
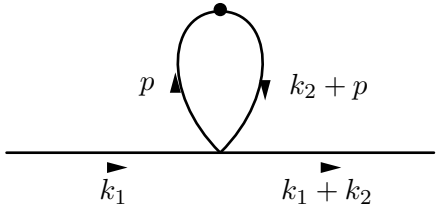

Bootstrap methods, which take \eqref{non-renorm1} and \eqref{non-renorm2} as input, are useful for improving the estimates of unprotected scaling dimensions. The study of these was initiated in \cite{fmn72} as well, which found the exponent
\begin{equation}
\Delta_{\phi^2} = \frac{d - \varepsilon}{2} + \frac{\varepsilon}{3} + \left [ \psi(1) - 2 \psi \left ( \frac{d}{4} \right ) + \psi \left ( \frac{d}{2} \right ) \right ] \left ( \frac{\varepsilon}{3} \right )^2 + O(\varepsilon^3) \; . \label{dim-eps}
\end{equation}
One quantity, which to our knowledge has not been computed yet, is the dimension of the leading spin-2 operator $T_{\mu\nu}$. We can use $\Delta_T$ to distinguish between different long-range Ising models since it gives a rough measure of how nonlocal a theory is. This makes it important for the bootstrap which is agnostic to microscopic parameters like $s$. Constructing the operator
\begin{equation}
T_{\mu\nu} = \left ( \phi \partial_\mu \partial_\nu \phi - \frac{1}{d} \delta_{\mu\nu} \phi \partial^2 \phi \right ) + y \left ( \partial_\mu \phi \partial_\nu \phi - \frac{1}{d} \delta_{\mu\nu} (\partial \phi)^2 \right ) \; , \label{t-def}
\end{equation}
we choose $y = -\frac{\Delta_\phi + 1}{\Delta_\phi}$ in order to make it a conformal primary.\footnote{It is easy to check that for $(\Delta_\phi, d) = (1, 4)$, \eqref{t-def} is the improved stress-energy tensor of a free scalar up to a constant factor.} The one-loop diagram that one might expect to give anomalous scaling to $\left < \phi(-k_1 - k_2) \phi(k_1) T_{\mu\nu}(k_2) \right >$ is shown in Figure \ref{graph-t1}.

After Feynman parameters are first invoked, the integral takes the form
\begin{eqnarray}
G(k_1, k_2) &=& \lambda \mu^\varepsilon \frac{\Gamma(s)}{\Gamma \left ( \frac{s}{2} \right )^2} \int_0^1 \int_{\mathbb{R}^d} \frac{x^{\frac{s}{2} - 1} (1 - x)^{\frac{s}{2} - 1}}{[(k_2 + p)^2 x + p^2 (1 - x)]^s} \label{1loop-t1} \\
&& \left [ (k_2 + p)_\nu ((k_2 + p)_\mu - yp_\mu) + p_\nu(p_\mu - y(k_2 + p)_\mu) - \mathrm{trace} \right ] \frac{\textup{d}p}{(2\pi)^d} \textup{d}x \; . \nonumber
\end{eqnarray}
After making the denominator spherically symmetric in $p$, the piece in brackets becomes $(p + (1 - x) k_2)_\nu [(p + (1 - x) k_2) - y (p - x k_2)_\mu] + (p - x k_2)_\nu [(p - x k_2)_\mu - y (p + (1 - x) k_2)_\mu]$ with the trace subtracted. Expanding this, all parts that are odd in a given component of $p_\mu$ must vanish. This reveals an overall factor of $k_{2\mu} k_{2\nu} - \frac{1}{d} \delta_{\mu\nu} k_2^2$ which we omit in what follows. The other parts of \eqref{1loop-t1} become
\begin{eqnarray}
G(k_1, k_2) &\propto& \frac{\lambda}{(4\pi)^{\frac{d}{2}}} \frac{\Gamma \left ( s - \frac{d}{2} \right )}{\Gamma \left ( \frac{s}{2} \right )^2} \left [ \frac{\Gamma \left ( \frac{d - s}{2} \right )^2}{\Gamma(d - s)} - 2(1 - y)\frac{\Gamma \left ( \frac{d - s}{2} + 1 \right )^2}{\Gamma(d - s + 2)} \right ] \left | \frac{k_2}{\mu} \right |^{-\varepsilon} \nonumber \\
&=& \frac{1}{\varepsilon} \frac{\lambda}{(4\pi)^{\frac{d}{2}} \Gamma \left ( \frac{d}{2} \right )} \left [ 2 - (1 - y) \frac{d}{d + 2} \right ] \left | \frac{k_2}{\mu} \right |^{-\varepsilon} + O(1) \; . \label{1loop-t2}
\end{eqnarray}
After inserting the leading order value of $y$, it becomes clear that \eqref{1loop-t2} is completely regular. This vanishing one-loop anomalous dimension, which had to be the case in $d = 4$, is actually true for all other values of $d$ as well.\footnote{This is in fact reassuring given that $k_{1\mu}k_{1\nu} + (k_1 + k_2)_\mu(k_1 + k_2)_\nu - yk_{1\mu}(k_1 + k_2)_\nu - yk_{1\nu}(k_1 + k_2)_\mu$ is the tree level contribution to $\left < \phi(-k_1 - k_2) \phi(k_1) T_{\mu\nu}(k_2) \right >$. This cannot be produced by Figure \ref{graph-t1} whose tensor structure only involves $k_2$.} A two-loop diagram is therefore necessary to see perturbative corrections in $\Delta_T$.
\begin{figure}[h]
\centering
\vspace{-2cm}
\begin{fmffile}{graph3}
\begin{fmfgraph*}(320, 200)
\fmfleft{p1}
\fmfright{p2}
\fmftop{t}
\fmf{plain}{p1,i1,i2,p2}
\fmf{plain,left}{i1,i2,i1}
\fmfforce{0.5w,0.615h}{t}
\fmfdot{t}
\fmffreeze

\arrowhack{a}{-4}{$q$}{10}{p1,p2}
\arrowhack{b}{-4}{$k_1$}{10}{p1,i1}
\arrowhack{c}{-4}{$k_1 + k_2$}{10}{i2,p2}
\arrowhack{d}{-21}{$k_1 - p - q$}{31}{p1,p2}
\arrowhack{e}{12}{$p$}{-32}{i1,t}
\arrowhack{f}{12}{$k_2 + p$}{-35}{t,i2}
\end{fmfgraph*}
\end{fmffile}
\vspace{-2cm}
\caption{The most interesting two-loop contribution to $\left < \phi\phi T \right >$.}
\label{graph-t2}
\end{figure}
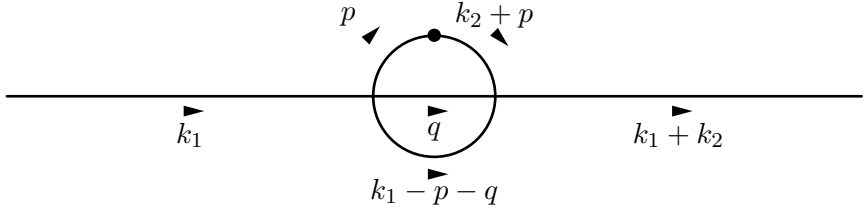

The two-loop $\gamma_T$ comes from the diagram in Figure \ref{graph-t2}. All other candidates either involve a scaleless loop or have Figure \ref{graph-t1} as a subdiagram. While standard methods are not enough to evaluate this integral in general, it is sufficient to set $k_1 = 0$ if we are only interested in the pole term. Starting with the $q$ integral already evaluated,
\begin{eqnarray}
G(0, k_2) &=& \frac{\lambda^2 \mu^{2\varepsilon}}{(4\pi)^{\frac{d}{2}}} \frac{\Gamma \left ( s - \frac{d}{2} \right ) \Gamma \left ( \frac{s - \varepsilon}{2} \right )^2}{\Gamma(s - \varepsilon) \Gamma \left ( \frac{s}{2} \right )^2} \int_{\mathbb{R}^d} \frac{1}{|p|^{s + \varepsilon} |k_2 + p|^s} \nonumber \\
&& \left [ (k_2 + p)_\nu ((k_2 + p)_\mu - yp_\mu) + p_\nu(p_\mu - y(k_2 + p)_\mu) - \mathrm{trace} \right ] \frac{\textup{d}p}{(2\pi)^d} \nonumber \\
&\propto& \frac{\lambda^2 \mu^{2\varepsilon}}{(4\pi)^{\frac{d}{2}}} \frac{\Gamma \left ( s - \frac{d}{2} \right ) \Gamma \left ( \frac{s - \varepsilon}{2} \right )^2 \Gamma \left ( s + \frac{\varepsilon}{2} \right )}{\Gamma(s - \varepsilon) \Gamma \left ( \frac{s}{2} \right )^3 \Gamma \left ( \frac{s + \varepsilon}{2} \right )} \nonumber \\
&& \int_0^1 \int_{\mathbb{R}^d} \frac{x^{\frac{s}{2} - 1} (1 - x)^{\frac{s + \varepsilon}{2} - 1}}{[p^2 + x (1 - x) k_2^2]^{s + \frac{\varepsilon}{2}}} [1 - 2x(1 - x)(1 - y)] \frac{\textup{d}p}{(2\pi)^d} \textup{d}x \; . \label{2loop-t1}
\end{eqnarray}
Once there is only a Feynman paremeter left to deal with, its integral can be expanded as a series in $\varepsilon$. The first term vanishes, as it must by consistency, but the second term does not. This gives the correlator an overall $\frac{1}{\varepsilon}$ dependence.\footnote{At $d = 4$, it is actually $\gamma_T + 2\gamma_\phi$ that appears in the Callan-Symanzik equation, explaining why \eqref{2loop-t2} still has a pole in this limit.}
\begin{eqnarray}
G(0, k_2) &\propto& \frac{\lambda^2}{(4\pi)^d} \frac{\Gamma \left ( s - \frac{d}{2} \right ) \Gamma \left ( \frac{s - \varepsilon}{2} \right )^2 \Gamma \left ( s + \frac{\varepsilon - d}{2} \right )}{\Gamma(s - \varepsilon) \Gamma \left ( \frac{s}{2} \right )^3 \Gamma \left ( \frac{s + \varepsilon}{2} \right )} \left | \frac{k_2}{\mu} \right |^{-2\varepsilon} \nonumber \\
&& \int_0^1 x^{\frac{s}{2} - \varepsilon - 1} (1 - x)^{\frac{s - \varepsilon}{2} - 1} [1 - 2x(1 - x)(1 - y)] \textup{d}x \nonumber \\
&=& \frac{1}{\varepsilon} \frac{\lambda^2}{(4\pi)^d \Gamma \left ( \frac{d}{2} \right )^2} \frac{4}{d(d + 2)} \left | \frac{k_2}{\mu} \right |^{-2\varepsilon} + O(1) \; . \label{2loop-t2}
\end{eqnarray}
From this expression, $\gamma_T$ can easily be read off. Substituting the fixed point, we have
\begin{equation}
\Delta_T = \frac{d + 4 - \varepsilon}{2} - \frac{8}{d(d + 2)} \left ( \frac{\varepsilon}{3} \right )^2 + O(\varepsilon^3) \; . \label{dim-t}
\end{equation}

\subsection{An infrared duality}
If one wishes to approximate critical exponents near values of $s$ where they are known exactly, the flow emanating from \eqref{flow1-action} only solves half of the problem. An ideal scenario would also explain how $\Delta_T$ and $\Delta_{\phi^2}$ approach $d$ and $\Delta_\epsilon^{\mathrm{SRI}}$ respectively. In \cite{brrz17a}, it was realized that the LRI at $s = s_*$ must contain more than just the SRI. Exact results in 2D and numerical results in 3D have established that the Ising model does not contain scalar primaries that form a shadow pair, nor does it have a vector that is able to recombine with $T_{\mu\nu}$. These problems were solved by introducing a generalized free scalar $\chi$ which plays the role of $\phi^3$. This also makes it possible to write down the (not unit-normalized) conformal primary
\begin{equation}
V_\mu = \Delta_\sigma \sigma \partial_\mu \chi - \Delta_\chi \chi \partial_\mu \sigma \; , \label{recombining-vector}
\end{equation}
which has the same dimension as $\partial^\nu T_{\mu\nu}$.

The newly introduced kinetic term can be tuned such that $\sigma \chi$ has dimension $d - \delta$ where $\delta \equiv \frac{s_* - s}{2}$. Viewing this operator as a deformation analogous to $\phi^4$, \cite{brrz17a} conjectured that
\begin{equation}
S = S_{\mathrm{SRI}} + \int -\frac{1}{2} \chi \partial^{-s} \chi + g\sigma\chi \textup{d}x \label{flow2-action}
\end{equation}
flows to the same LRI fixed point as \eqref{flow1-action}. The statement of \eqref{flow2-action} is that one can deform an exactly known correlation function by inserting $\exp \left ( g \int \sigma\chi \textup{d}x \right )$ and applying the rules of conformal perturbation theory \cite{z87}. Encouraging results also follow from deforming correlators that are known with high precision from the numerical bootstrap \cite{ks17}. Reviewing the results of \cite{brrz17a, brrz17b}, the beta function is
\begin{equation}
\beta(g) = \begin{cases}
-\delta g + 1.268404 g^3 + O(g^5) & d = 2 \\
-\delta g + 12.26(3) g^3 + O(g^5) & d = 3
\end{cases} \label{flow2-beta}
\end{equation}
and the system in the IR preserves the relations \eqref{non-renorm1} and \eqref{non-renorm2}. The expansion dual to \eqref{dim-eps} is
\begin{equation}
\Delta_\epsilon = \begin{cases}
1 + O(g^4) & d = 2 \\
1.412625(10) + 3.3(5) g^2 + O(g^4) & d = 3
\end{cases} \label{dim-eps2}
\end{equation}
and the one dual to \eqref{dim-t} is
\begin{equation}
\Delta_T = \begin{cases}
2 + \frac{15}{32} \pi^2 g^2 + O(g^4) & d = 2 \\
3 + 28.60555(6) g^2 + O(g^4) & d = 3
\end{cases} \; . \label{dim-t2}
\end{equation}
Some expressions in terms of $g$ are exact, which is why we have refrained from plugging in the fixed point. The reason why $\Delta_T$ is known much more precisely than $\Delta_\epsilon$ in three dimensions is that anomalous dimensions of broken currents must be compatible with multiplet recombination. Specifically, the formula in \cite{gk16} allows \eqref{dim-t2} to be calculated purely from the contact term in the three-point function $\left < T_{\mu\nu} V_\rho \sigma \chi \right >$. From now on, we will use the $(\phi, \phi^2, \phi^3)$ and $(\sigma, \epsilon, \chi)$ notation interchangeably.

\subsection{Exact OPE coefficient ratios}
In addition to the non-renormalization theorems we have discussed, the nonlocal equation of motion can be used to derive an infinite family of ratios between OPE coefficients. Equations like \eqref{nonlocal-eom} are statements about fields in a Lagrangian rather than unit-normalized operators in a CFT. We will therefore equip them with $s$-dependent prefactors which we call $n_\sigma$ and $n_\chi$. Writing the integral expression for \eqref{nonlocal-eom} yields
\begin{equation}
n_\chi(s) \chi(x) = \int \frac{n_\sigma(s) \sigma(y)}{|x - y|^{d + s}} \textup{d}y \; , \label{better-eom}
\end{equation}
which is recognizable as the shadow transform. The idea is to use \eqref{better-eom} in three-point functions containing $\sigma$ and $\chi$ in order to find
\begin{equation}
\frac{\lambda_{12\chi}}{\lambda_{12\sigma}} = \frac{n_\sigma(s)}{n_\chi(s)} R_{12} \; , \label{ratio-schematic}
\end{equation}
where $R_{12}$ is a known function of the quantum numbers. If we repeat this for a second pair of operators, we can arrange to have the normalizations cancel out, leaving us with
\begin{equation}
\frac{\lambda_{12\chi} \lambda_{34\sigma}}{\lambda_{12\sigma} \lambda_{34\chi}} = \frac{R_{12}}{R_{34}} \; . \label{ratio-cancelled}
\end{equation}
The explicit $R_{ij}$ was computed for scalars in \cite{prvz16}. Subsequently, \cite{brrz17a} used it to test the IR duality between the flows that generate the $\varepsilon$-expansion and the $\delta$-expansion. In what follows, we aim to generalize this result to the case of spinning operators.

A natural language for this is the embedding formalism \cite{d36} which associates to each $x^\mu \in \mathbb{R}^d$ a null ray $X^M \sim \lambda X^M \in \mathbb{R}^{d + 1, 1}$. If one chooses representatives such that $X^+ = 1$ (the Poincar\'e section of the null cone), a Lorentz transformation on $X^M = (1, x^2, x^\mu)$ precisely implements a conformal transformation on $x^\mu$. Conformally invariant quantities can therefore be built out of the Lorentz scalars
\begin{equation}
X_{ij} \equiv -2 X_i \cdot X_j = x_{ij}^2 \; . \label{dot-product}
\end{equation}
An especially useful incarnation of the embedding space was developed in \cite{cppr11a, cppr11b} which used polarization vectors to make the formalism index-free. One such vector $Z$, in addition to being null, must satisfy a transversaity condition with $X$:
\begin{equation}
X_{ii} = 0 \; , \; X_i \cdot Z_i = 0 \; , \; Z_{ii} = 0 \; . \label{drop-these}
\end{equation}
This is because tracelessness in $\mathbb{R}^{d + 1, 1}$ is stronger than tracelessness in $\mathbb{R}^d$.
An important result is that correlation functions may only depend on polarization vectors through the combinations
\begin{eqnarray}
H_{ij} &\equiv& -2 [(Z_i \cdot Z_j)(X_i \cdot X_j) - (X_1 \cdot Z_2)(X_2 \cdot Z_1)] \nonumber \\
V_{i, jk} &\equiv& \frac{(Z_i \cdot X_j)X_{ik} - (Z_i \cdot X_k)X_{ij}}{X_{jk}} \; . \label{hv-structures}
\end{eqnarray}

We will begin with the three-point function $\left < \phi_1(x_1) \mathcal{O}_2^{\mu_1, \dots, \mu_\ell}(x_2) \sigma(x_3) \right >$, which has a single tensor structure. The most straightforward extension of it to the projective null cone is
\begin{equation}
\left < \Phi_1(X_1) \mathcal{O}_2(X_2, Z_2) \sigma(X_3) \right > = \frac{\lambda_{12\sigma} V_{2,13}^\ell}{X_{13}^{\frac{\Delta_\sigma + \Delta_{12} - \ell}{2}} X_{23}^{\frac{\Delta_\sigma - \Delta_{12} + \ell}{2}} X_{12}^{\frac{\Delta_1 + \Delta_2 - \Delta_\sigma + \ell}{2}}} \; . \label{lifted-3pt}
\end{equation}
As a check, this is a degree-$\ell$ polynomial in $Z_2$ which is invariant under $Z_2 \mapsto Z_2 + X_2$. It also transforms with the correct weights when $X_1$, $X_2$ and $X_3$ are scaled individually. Lifting the equation of motion to the embedding space as well,
\begin{equation}
\left < \Phi_1(X_1) \mathcal{O}_2(X_2, Z_2) \chi(X_3) \right > = \frac{n_\sigma(s)}{n_\chi(s)} \frac{\lambda_{12\sigma}}{X_{12}^{\frac{\Delta_1 + \Delta_2 - \Delta_\sigma + \ell}{2}}} \int \frac{V_{2,10}^\ell DX_0}{X_{01}^{\frac{\Delta_\sigma + \Delta_{12} - \ell}{2}} X_{02}^{\frac{\Delta_\sigma - \Delta_{12} + \ell}{2}} X_{03}^{d - \Delta_\sigma}} \; . \label{cintegral1}
\end{equation}
This type of object, which has exponents adding up to $d$, is called a conformal integral. Suitable technology for treating conformal integrals in the embedding space, including the formula
\begin{equation}
\int \frac{X^{A_1} \dots X^{A_n}}{(-2 X \cdot Y)^{d + n}} DX = \frac{\pi^{\frac{d}{2}} \Gamma \left ( \frac{d}{2} + n \right )}{\Gamma(d + n)} \frac{Y^{A_1} \dots Y^{A_n}}{(-Y^2)^{\frac{d}{2} + n}} - \mathrm{traces} \; , \label{cintegral-trick}
\end{equation}
was developed in \cite{s14}. For similar integrals with a slight excess in the exponents, see \cite{st17}. Before returning to \eqref{cintegral1}, it is worth expanding the tensor structure as
\begin{equation}
V_{2,10}^\ell = X_{01}^{-\ell} \sum_{n = 0}^{\infty} \binom{\ell}{n} (Z_2 \cdot X_1)^n (Z_2 \cdot X_0)^{\ell - n} X_{02}^n X_{12}^{\ell - n} \; . \label{v-expansion}
\end{equation}
The result of \eqref{cintegral1} will contain $V_{2,13}^\ell$ and in particular $(Z_2 \cdot X_3)^\ell$, which can only come from the first term of \eqref{v-expansion}. It is therefore enough to focus on the $n = 0$ term and infer the others from conformal invariance. Introducing Schwinger parameters and using \eqref{cintegral-trick}, we have
\begin{eqnarray}
\left < \Phi_1(X_1) \mathcal{O}_2(X_2, Z_2) \chi(X_3) \right > &=& \frac{n_\sigma(s)}{n_\chi(s)} \frac{\lambda_{12\sigma} Z_2^{A_1} \dots Z_2^{A_\ell}}{X_{12}^{\frac{\Delta_1 + \Delta_2 - \Delta_\sigma - \ell}{2}}} \frac{\pi^{\frac{d}{2}} \Gamma \left ( \frac{d}{2} + \ell \right )}{\Gamma(d - \Delta_\sigma) \Gamma \left ( \frac{\Delta_\sigma + \Delta_{12} + \ell}{2} \right ) \Gamma \left ( \frac{\Delta_\sigma - \Delta_{12} + \ell}{2} \right )} \nonumber \\
&& \int_0^\infty \int_0^\infty \frac{(X_3 + \alpha X_1 + \beta X_2)^{A_1} \dots (X_3 + \alpha X_1 + \beta X_2)^{A_\ell}}{\alpha^{-\frac{\Delta_\sigma + \Delta_{12} + \ell}{2}} \beta^{-\frac{\Delta_\sigma - \Delta_{12} + \ell}{2}} [\alpha X_{13} + \beta X_{23} + \alpha \beta X_{12}]^{\frac{d}{2} + \ell}} \frac{\textup{d}\alpha}{\alpha} \frac{\textup{d}\beta}{\beta} \nonumber \\
&& + O(Z_2 \cdot X_1) \; . \label{cintegral2}
\end{eqnarray}
If we again discard $Z_2 \cdot X_1$ terms, we can evaluate the integral to arrive at
\begin{equation}
R_{12} = \pi^{\frac{d}{2}} \frac{\Gamma \left ( \Delta_\sigma - \frac{d}{2} \right ) \Gamma \left ( \frac{d - \Delta_\sigma + \Delta_{12} + \ell}{2} \right ) \Gamma \left ( \frac{d - \Delta_\sigma - \Delta_{12} + \ell}{2} \right )}{\Gamma(d - \Delta_\sigma) \Gamma \left ( \frac{\Delta_\sigma + \Delta_{12} + \ell}{2} \right ) \Gamma \left ( \frac{\Delta_\sigma - \Delta_{12} + \ell}{2} \right )} \; . \label{ratio}
\end{equation}

This logic can be repeated for correlators that have arbitrary spin in both positions. The difference here is that there is no longer a unique tensor structure:
\begin{equation}
\left < \mathcal{O}_1(X_1, Z_1) \mathcal{O}_2(X_2, Z_2) \sigma(X_3) \right > = \sum_{m = 0}^{\mathrm{min}(\ell_1, \ell_2)} \lambda_{12\sigma}^{(m)} \frac{V_{1,23}^{\ell_1 - m} V_{2,13}^{\ell_2 - m} H_{12}^m}{X_{13}^{\frac{\Delta_\sigma + \tau_{12}}{2}} X_{23}^{\frac{\Delta_\sigma - \tau_{12}}{2}} X_{12}^{\frac{\tau_1 + \tau_2 - \Delta_\sigma}{2}}} \; . \label{lifted-general}
\end{equation}
However, the factor of $H_{12}^m$ is untouched by the integration. This means that $\lambda_{12\chi}^{(m)}$ is proportional to $\lambda_{12\sigma}^{(m)}$ and $R_{ij}$ acquires one extra index instead of two. Carrying out the computation, we find
\begin{equation}
R_{12}^{(m)} = \pi^{\frac{d}{2}} \frac{\Gamma \left ( \Delta_\sigma - \frac{d}{2} \right ) \Gamma \left ( \frac{d - \Delta_\sigma + \Delta_{12} + \ell_1 + \ell_2 - 2m}{2} \right ) \Gamma \left ( \frac{d - \Delta_\sigma - \Delta_{12} + \ell_1 + \ell_2 - 2m}{2} \right )}{\Gamma(d - \Delta_\sigma) \Gamma \left ( \frac{\Delta_\sigma + \Delta_{12} + \ell_1 + \ell_2 - 2m}{2} \right ) \Gamma \left ( \frac{\Delta_\sigma - \Delta_{12} + \ell_1 + \ell_2 - 2m}{2} \right )} \; . \label{ratio-general}
\end{equation}
As a further generalization, one could consider mixed-symmetry tensors using the formalism of \cite{ch15}.

\section{Exploiting the shadow relation}
\label{sec:shadow}
The conformal block expansion of $\left < \sigma\sigma\chi\chi \right >$ is heavily constrained by the OPE coefficient ratio \eqref{ratio}. Choosing three-point functions that involve the two shadow operators and a traceless symmetric primary $\mathcal{O}$, the quadratic equality
\begin{equation}
\lambda^2_{\sigma\chi\mathcal{O}} = \frac{R_{\sigma\mathcal{O}}}{R_{\chi\mathcal{O}}} \lambda_{\sigma\sigma\mathcal{O}}\lambda_{\chi\chi\mathcal{O}} \label{super-ratio}
\end{equation}
immediately follows. Two results that follow from \eqref{super-ratio} are worth fleshing out in detail as both of them provide useful input to the numerical bootstrap.

\subsection{A tower of protected operators}
We will first consider a primary $\mathcal{O}$ whose spin is odd. In this case, the OPE coefficients on the right hand side of \eqref{super-ratio} vanish by Bose symmetry. For the left hand side to be nonzero, the dimension of $\mathcal{O}$ must be a pole of the following expression.
\begin{equation}
\frac{R_{\sigma\mathcal{O}}}{R_{\chi\mathcal{O}}} = \frac{\Gamma \left ( \frac{d - \Delta + \ell}{2} \right )^2 \Gamma \left ( \frac{d - 2\Delta_\sigma + \Delta + \ell}{2} \right ) \Gamma \left ( \frac{2\Delta_\sigma - d + \Delta + \ell}{2} \right )}{\Gamma \left ( \frac{\Delta + \ell}{2} \right )^2 \Gamma \left ( \frac{2\Delta_\sigma - \Delta + \ell}{2} \right ) \Gamma \left ( \frac{2d - 2\Delta_\sigma - \Delta + \ell}{2} \right )} \label{super-ratio-expression}
\end{equation}
The poles above the unitarity bound, which come entirely from the squared gamma function in the numerator, are
\begin{equation}
\Delta = d + \ell + 2n = \Delta_\sigma + \Delta_\chi + \ell + 2n \; . \label{super-ratio-poles}
\end{equation}
These are nothing but the dimensions of the double-twist operators $[\sigma\chi]_{n, \ell} \sim \chi \partial_{\mu_1} \dots \partial_{\mu_\ell} \partial^{2n} \sigma$ at $s = s_*$. The first element of this list, $[\sigma\chi]_{0, 1}$, is known to renormalize upon lowering the value of $s$. As its dimension leaves the pole \eqref{super-ratio-poles}, the left hand side of \eqref{super-ratio} is able to remain nonzero. This is because $[\sigma\chi]_{0,1}$ recombines with the stress-energy tensor and Bose symmetric OPEs are allowed to contain odd-spin descendants. All of the other double-twist operators stay primary (at least in three dimensions) and thus face a radically different situation. Bose symmetry continues to enforce $\lambda_{\sigma\sigma\mathcal{O}} = \lambda_{\chi\chi\mathcal{O}} = 0$ which means that $\Delta$ can only change continuously if $\lambda^2_{\sigma\chi\mathcal{O}}$ jumps to zero discontinuously. We therefore arrive at the following proposal.

\vspace{0.3cm}
\textit{In the long-range Ising model given by a generic $\frac{d}{2} < s < s_*$, all odd-spin primaries in $\sigma \times \chi$ (the double-twist operators other than the first one) have a scaling dimension that is independent of $s$.}

\begin{figure}[h]
\centering
\includegraphics[scale=0.45]{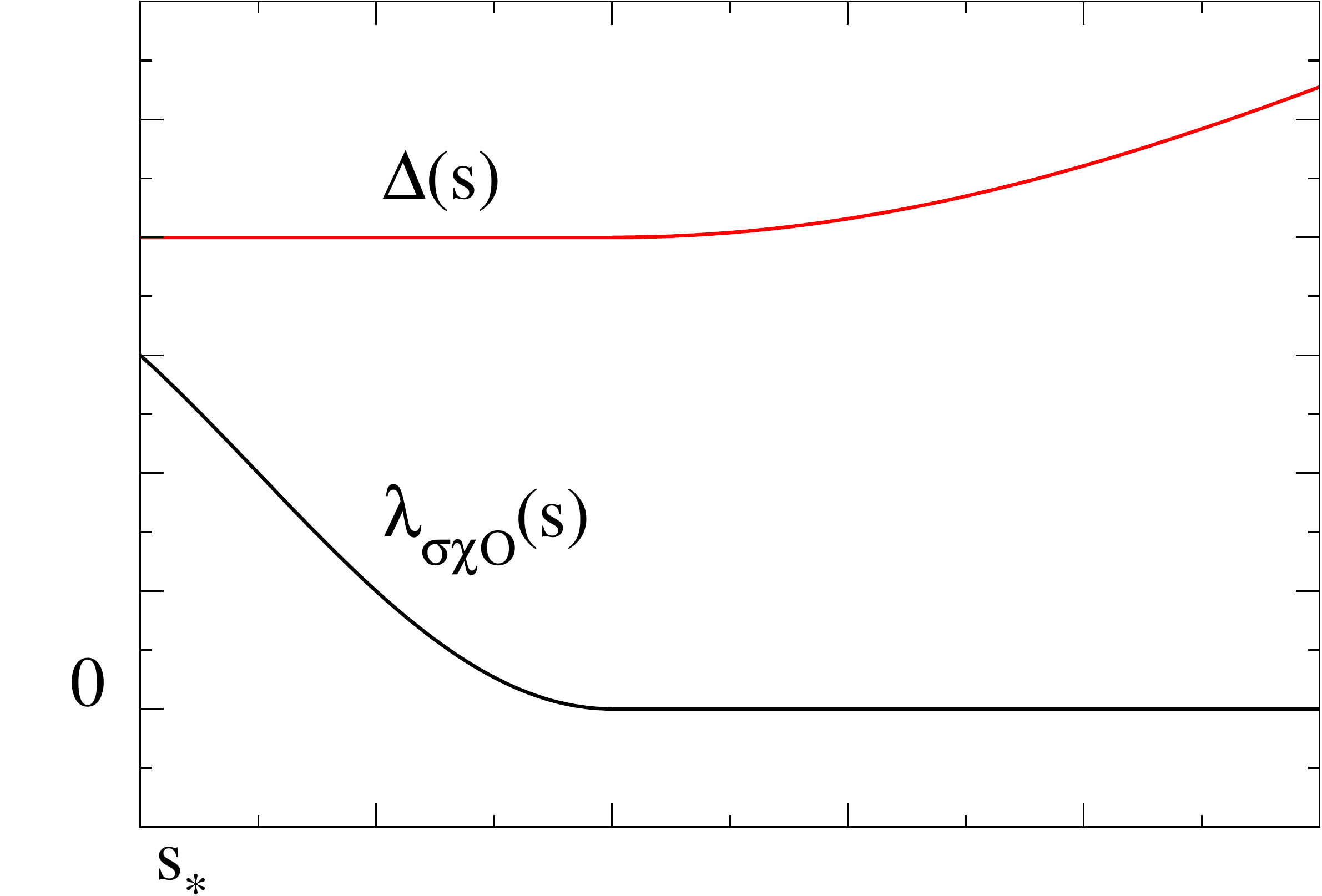}
\caption{A possible way out of our non-renormalization theorem for odd-spin primary operators in $\sigma \times \chi$. The strongest statement we can make is that their dimensions are protected in an open neighbourhood of any point where $\lambda^2_{\sigma\chi\mathcal{O}} > 0$. It could be the case that at least one renormalizes after decoupling at an intermediate value of $s$.}
\label{fig:smooth-approach}
\end{figure}

This result, which should strictly be called a conjecture, would represent the most natural scenario even if we did not have continuity in $s$. It also follows from the earlier form of the OPE ratio \eqref{ratio-schematic} after a simple check that the normalizations $n_\sigma(s)$ and $n_\chi(s)$ are nonzero. One possibility that we cannot rule out is that these odd-spin primaries are only protected within a finite interval starting at $s = s_*$. This is because we used the fact that $\lambda^2_{\sigma\chi\mathcal{O}}$ was strictly positive. If this coefficient were to smoothly approach zero at some value of $s$, we would have to worry about the behaviour in Figure \ref{fig:smooth-approach}.

Running this argument around the other side of the duality is more subtle since $\phi$ and $\phi^3$ cannot be treated as independent fields. At small values of the spin, however, it is clear how the protected operators in $\phi \times \phi^3$ match up with those in $\sigma \times \chi$. Taking $n = 0$ for example,
\begin{eqnarray}
T_{\mu\nu} \ni \chi \partial_\mu \sigma & & \phi^3 \partial_\mu \phi \in \phi^4 \nonumber \\
\chi \partial_\mu \partial_\nu \partial_\rho \sigma & \leftrightarrow & \phi^3 \partial_\mu \partial_\nu \partial_\rho \phi \nonumber \\
\chi \partial_\mu \partial_\nu \partial_\rho \partial_\sigma \partial_\tau \sigma & \leftrightarrow & \phi^3 \partial_\mu \partial_\nu \partial_\rho \partial_\sigma \partial_\tau \phi \nonumber \\
&\dots& \label{initial-matching}
\end{eqnarray}
Starting at $\ell = 7$, the conformal primary where four copies of $\phi$ are saturated by derivatives is not unique. A generating function counting the number of such primaries for all $\ell$ can be found in \cite{r16b}. Looking at one example, the primary subspace for operators of the form
\begin{eqnarray}
\mathcal{O}_7 &=& c_0 \phi^3 \partial^7 \phi + c_1 \phi^2 \partial \phi \partial^6 \phi + c_2 \phi^2 \partial^2 \phi \partial^5 \phi + c_3 \phi \partial \phi \partial \phi \partial^5 \phi \nonumber \\
&& + c_4 \phi^2 \partial^3 \phi \partial^4 \phi  + c_5 \phi \partial \phi \partial^2 \phi \partial^4 \phi + c_6 \partial \phi \partial \phi \partial \phi \partial^4 \phi + c_7 \phi \partial \phi \partial^3 \phi \partial^3 \phi \nonumber \\
&& + c_8 \phi \partial^2 \phi \partial^2 \phi \partial^3 \phi + c_9 \partial \phi \partial \phi \partial^2 \phi \partial^3 \phi + c_{10} \partial \phi \partial^2 \phi \partial^2 \phi \partial^2 \phi \label{ansatz-7}
\end{eqnarray}
is two-dimensional. We should therefore only expect one linear combination to be protected. Given a basis consisting of $\left \{ \mathcal{O}_7^{(1)}, \mathcal{O}_7^{(2)} \right \}$, we may choose $t$ such that $t \mathcal{O}_7^{(1)} + \mathcal{O}_7^{(2)}$ decouples from $\phi \times \phi^3$. This gives an operator that is free to renormalize since it has vanishing OPE coefficients on either side of \eqref{super-ratio}. It is only the orthogonal operator that maintains its exact double-twist dimension. More generally, when the subspace is $N$-dimensional, the solutions to
\begin{equation}
t_1 \left < \phi(x_1) \phi^3(x_2) \mathcal{O}_\ell^{(1)} \right > + \dots + t_{N - 1} \left < \phi(x_1) \phi^3(x_2) \mathcal{O}_\ell^{(N - 1)} \right > + \left < \phi(x_1) \phi^3(x_2) \mathcal{O}_\ell^{(N)} \right > = 0 \label{decoupling-subspace}
\end{equation}
are $(N - 1)$-dimensional, pointing us to a unique protected operator once again. It is not surprising that $\phi \times \phi^3$ contains only one leading-twist operator of each spin in a suitable basis. Indeed if the degeneracy could not be removed, one would be able to repeat our non-renormalization argument based on the nonlocal equation of motion $\partial^s \phi \propto \phi^3$ in the local fixed point governed by $\partial^2 \phi \propto \phi^3$. It is clear that the Wilson-Fisher fixed point does not have an odd-spin protected tower. Instead, the odd-spin operators which have the dimensions \eqref{super-ratio-poles} in $d = 4$ are able to avoid the Bose symmetry constraints for primaries by recombining with higher spin currents. It would be interesting to see how this structure is reproduced in perturbation theory.

The 2D case would also be very interesting to study further. This describes a fixed line obtained by deforming a Virasoro symmetric theory. As a result, the definition of $[\sigma\chi]_{n, \ell}$ is ambiguous just like the double-twist operator $[\phi\phi^3]_{n, \ell}$. This time, rotating the bases to remove the maximal number of operators from $\sigma \times \chi$ is probably not the right solution to the mixing problem. Consider the operators with $(h, \bar{h}) = (4, 1)$ where
\begin{eqnarray}
\mathcal{O}_3^{(1)} &=& \sigma L_{-1}^3 \chi - 93 L_{-1} \sigma L_{-1}^2 \chi + \frac{713}{3} L_{-1}^2 \sigma L_{-1} \chi - \frac{3565}{51} L_{-1}^3 \sigma \chi \nonumber \\
\mathcal{O}_3^{(2)} &=& L_{-3} \sigma \chi - \frac{128}{153} L_{-1}^3 \sigma \chi \label{ansatz-3}
\end{eqnarray}
is a valid basis.\footnote{We have implicitly used the null state condition to write $L_{-2} \sigma$ as a multiple of $L_{-1}^2 \sigma$. Similarly, we do not have any higher Virasoro generators acting on $\chi$ since the theory it comes from is already nonlocal at $s = s_*$.} The previous logic would suggest splitting these into one combination with $\lambda_{\sigma\chi\mathcal{O}} = 0$ and another combination with $\lambda_{\sigma\chi\mathcal{O}} \neq 0$. This presents a problem as both operators in such a splitting would have nonzero overlap with
\begin{equation}
\Lambda = \left ( L_{-4} - \frac{5}{3} L_{-2}^2 \right ) I \; . \label{current-4}
\end{equation}
There is only room for a higher spin current to recombine with one multiplet so we must demand that the protected spin-3 operator is the one that fails to give an anomalous dimension to \eqref{current-4}. To solve for this operator, we have evaluated the three-point functions
\begin{eqnarray}
\left < \Lambda(z_1, \bar{z}_1) \mathcal{O}_3^{(1)}(z_2, \bar{z}_2) \sigma\chi(z_3, \bar{z}_3) \right > &=& \frac{124775 / 544}{z_{12}^7 z_{13} z_{23} \bar{z}_{23}^2} \nonumber \\
\left < \Lambda(z_1, \bar{z}_1) \mathcal{O}_3^{(2)}(z_2, \bar{z}_2) \sigma\chi(z_3, \bar{z}_3) \right > &=& \frac{1225 / 3264}{z_{12}^7 z_{13} z_{23} \bar{z}_{23}^2} \label{ward-id-correlators}
\end{eqnarray}
using the Virasoro Ward identity.\footnote{The form $\left < L_{-m}I(z_1) L_{-n}\sigma(z_2) \sigma(z_3) \right > = \oint_{z_1} \oint_{z_2} (y_1 - z_1)^{1 - m} (y_2 - z_2)^{1 - n} \left < T(y_1) T(y_2) \sigma(z_2) \sigma(z_3) \right > \frac{\textup{d}y_2}{2\pi i} \frac{\textup{d}y_1}{2\pi i}$ is the one most useful for our purposes.} This allows us to repeat the calculation that \cite{brrz17a} did for the stress-energy tensor and say that
\begin{eqnarray}
&& \int \left < \bar{\partial}_1 \Lambda(z_1, \bar{z}_1) \left [ \frac{544}{124775} \mathcal{O}_3^{(1)} - \frac{3264}{1225} \mathcal{O}_3^{(2)} \right ](z_2, \bar{z}_2) \sigma\chi(z_3, \bar{z}_3) \right > \textup{d}z_3 \textup{d}\bar{z}_3 = 0 \nonumber \\
&& \int \left < \bar{\partial}_1 \Lambda(z_1, \bar{z}_1) \left [ \frac{713}{1513} \mathcal{O}_3^{(1)} + \frac{3422400}{10591} \mathcal{O}_3^{(2)} \right ](z_2, \bar{z}_2) \sigma\chi(z_3, \bar{z}_3) \right > \textup{d}z_3 \textup{d}\bar{z}_3 \propto \gamma_\Lambda \; . \label{hs-recombination}
\end{eqnarray}
Based on this, one might hope that all $\sigma \times \chi$ operators in the 2D theory are either protected or eaten. The first counter-example to this appears at the next level which has three $\ell = 5$ operators built from $\sigma$ and $\chi$ but only one $\ell = 6$ current that needs to be broken.

\subsection{Consequences for crossing}
We will now discuss the treatment of $[\sigma\chi]_{n,\ell}$ operators with even spin. Looking at \eqref{super-ratio} for $s = s_*$, we again have a removable singularity since $\mathcal{O}$ has a dimension given by \eqref{super-ratio-poles} while decoupling from the $\sigma \times \sigma$ and $\chi \times \chi$ OPEs. The difference is that the coefficients $\lambda_{\sigma\sigma\mathcal{O}}$ and $\lambda_{\chi\chi\mathcal{O}}$ turn on for $s < s_*$ as they are not constrained by any kinematic principle. This allows the right hand side of \eqref{super-ratio} to become a ratio of finite numbers. The shadow relation then becomes a statement about operators of an unknown dimension still having constrained OPE coefficients.

This situation is ubiquitous in the superconformal bootstrap. It allows four-point functions to be decomposed into blocks that include the contributions of many conformal primaries. These superconformal blocks have been computed in \cite{do02} and many subsequent works. In the long-range Ising model which is non-supersymmetric, it is the nonlocal operator in \eqref{nonlocal-eom} rather than a supercharge, which allows certain conformal blocks to be combined.

To make this precise, consider the general form of the four-point function for scalar primaries
\begin{equation}
\left < \phi_i(x_1) \phi_j(x_2) \phi_k(x_3) \phi_l(x_4) \right > = \left ( \frac{|x_{24}|}{|x_{14}|} \right )^{\Delta_{ij}} \left ( \frac{|x_{14}|}{|x_{13}|} \right )^{\Delta_{kl}} \frac{G^{ijkl}(u, v)}{|x_{12}|^{\Delta_i + \Delta_j}|x_{34}|^{\Delta_k + \Delta_l}} \; . \label{4pt-form}
\end{equation}
The unknown function, depending on the cross-ratios $u = \frac{x_{12}^2 x_{34}^2}{x_{13}^2 x_{24}^2}$ and $v = \frac{x_{14}^2 x_{23}^2}{x_{13}^2 x_{24}^2}$, has the conformal block expansion
\begin{equation}
G^{ijkl}(u, v) = \sum_{\mathcal{O}} \lambda_{ij\mathcal{O}} \lambda_{kl\mathcal{O}} G^{\Delta_{ij} , \Delta_{kl}}_{\mathcal{O}}(u, v) \; . \label{block-sum-form}
\end{equation}
By demanding crossing symmetry for the $\left < \sigma\sigma\chi\chi \right >$ correlator, we derive the crossing equations
\begin{eqnarray}
&& \sum_{\mathcal{O}^+_{2 | \ell}} \lambda_{\sigma\sigma\mathcal{O}} \lambda_{\chi\chi\mathcal{O}} F^{\sigma\sigma ; \chi\chi}_{-,\Delta,\ell}(u, v) + \sum_{\mathcal{O}^+_{2 | \ell}} \lambda_{\sigma\chi\mathcal{O}}^2 F^{\chi\sigma ; \sigma\chi}_{-,\Delta,\ell}(u, v) - \sum_{\mathcal{O}^+_{2 \nmid \ell}} \lambda_{\sigma\chi\mathcal{O}}^2 F^{\chi\sigma ; \sigma\chi}_{-,\Delta,\ell}(u, v) = 0 \label{change0} \\
&& \sum_{\mathcal{O}^+_{2 | \ell}} \lambda_{\sigma\sigma\mathcal{O}} \lambda_{\chi\chi\mathcal{O}} F^{\sigma\sigma ; \chi\chi}_{+,\Delta,\ell}(u, v) - \sum_{\mathcal{O}^+_{2 | \ell}} \lambda_{\sigma\chi\mathcal{O}}^2 F^{\chi\sigma ; \sigma\chi}_{+,\Delta,\ell}(u, v) + \sum_{\mathcal{O}^+_{2 \nmid \ell}} \lambda_{\sigma\chi\mathcal{O}}^2 F^{\chi\sigma ; \sigma\chi}_{+,\Delta,\ell}(u, v) = 0 \; , \nonumber
\end{eqnarray}
where we have used a shorthand for the convolved conformal block
\begin{equation}
F_{\pm, \Delta, \ell}^{ij ; kl}(u, v) = v^{\frac{\Delta_j + \Delta_k}{2}} G_{\Delta, \ell}^{\Delta_{ij}, \Delta_{kl}}(u, v) \pm u^{\frac{\Delta_j + \Delta_k}{2}} G_{\Delta, \ell}^{\Delta_{ij}, \Delta_{kl}}(v, u) \; . \label{convolved-block}
\end{equation}
We can modify \eqref{change0} to account for the protected operators
\begin{eqnarray}
&& \sum_{\mathcal{O}^+_{2 | \ell}} \lambda_{\sigma\sigma\mathcal{O}} \lambda_{\chi\chi\mathcal{O}} F^{\sigma\sigma ; \chi\chi}_{-,\Delta,\ell}(u, v) + \sum_{\mathcal{O}^+_{2 | \ell}} \lambda_{\sigma\chi\mathcal{O}}^2 F^{\chi\sigma ; \sigma\chi}_{-,\Delta,\ell}(u, v) - \sum_{\ell = 1, 3, \dots} \sum_{n = 0}^\infty \lambda_{\sigma\chi\mathcal{O}}^2 F^{\chi\sigma ; \sigma\chi}_{-,d + \ell + 2n,\ell}(u, v) = 0 \nonumber \\
&& \sum_{\mathcal{O}^+_{2 | \ell}} \lambda_{\sigma\sigma\mathcal{O}} \lambda_{\chi\chi\mathcal{O}} F^{\sigma\sigma ; \chi\chi}_{+,\Delta,\ell}(u, v) - \sum_{\mathcal{O}^+_{2 | \ell}} \lambda_{\sigma\chi\mathcal{O}}^2 F^{\chi\sigma ; \sigma\chi}_{+,\Delta,\ell}(u, v) + \sum_{\ell = 1, 3, \dots} \sum_{n = 0}^\infty \lambda_{\sigma\chi\mathcal{O}}^2 F^{\chi\sigma ; \sigma\chi}_{+,d + \ell + 2n,\ell}(u, v) = 0 \; , \nonumber \\
&& \label{change1}
\end{eqnarray}
but this only imposes the odd-spin case of the shadow relation. Imposing the even-spin case as well leads to
\begin{eqnarray}
&& \sum_{\mathcal{O}^+_{2 | \ell}} \lambda_{\sigma\sigma\mathcal{O}} \lambda_{\chi\chi\mathcal{O}} \mathcal{F}_{-,\Delta,\ell}(u, v) - \sum_{\ell = 1, 3, \dots} \sum_{n = 0}^\infty \lambda_{\sigma\chi\mathcal{O}}^2 F^{\chi\sigma ; \sigma\chi}_{-,d + \ell + 2n,\ell}(u, v) = 0 \nonumber \\
&& \sum_{\mathcal{O}^+_{2 | \ell}} \lambda_{\sigma\sigma\mathcal{O}} \lambda_{\chi\chi\mathcal{O}} \mathcal{F}_{+,\Delta,\ell}(u, v) + \sum_{\ell = 1, 3, \dots} \sum_{n = 0}^\infty \lambda_{\sigma\chi\mathcal{O}}^2 F^{\chi\sigma ; \sigma\chi}_{+,d + \ell + 2n,\ell}(u, v) = 0 \; , \label{change2}
\end{eqnarray}
where we have defined the convolved superblocks
\begin{equation}
\mathcal{F}_{\pm,\mathcal{O}}(u, v) = F^{\sigma\sigma ; \chi\chi}_{\pm,\Delta,\ell}(u, v) \mp \frac{R_{\sigma\mathcal{O}}}{R_{\chi\mathcal{O}}} F^{\chi\sigma ; \sigma\chi}_{\pm,\Delta,\ell}(u, v) \; . \label{superblocks}
\end{equation}
It is easy to read off what the non-convolved superblocks are.

In contrast to other known superblocks, \textit{e.g.} the 4D $\mathcal{N} = 1$ classification in \cite{ls16}, the relative coefficient in \eqref{superblocks} is not a rational function of $\Delta$. Indeed \eqref{super-ratio-expression} has infinitely many poles. It may therefore be of interest to compute rational approximations for the coefficient, similar to what is already standard practice for the conformal blocks themselves. We discuss both problems together in Appendix \ref{sec:appa}.

\section{Numerical results}
\label{sec:numerics}
We will now combine the numerical bootstrap with the exact results of the previous sections. We have already written two of the crossing equations which take the form \eqref{change0}, \eqref{change1} or \eqref{change2} depending on how much non-perturbative information is imposed. These are part of a larger system, given in Appendix \ref{sec:appa}, which has the schematic form
\begin{equation}
\sum_{\mathcal{O}^+_{2 | \ell}} \left ( \lambda_{\sigma\sigma\mathcal{O}} \; \lambda_{\epsilon\epsilon\mathcal{O}} \; \lambda_{\chi\chi\mathcal{O}} \right ) V^{(0)}_{\Delta, \ell} \left ( \begin{tabular}{c} $\lambda_{\sigma\sigma\mathcal{O}}$ \\ $\lambda_{\epsilon\epsilon\mathcal{O}}$ \\ $\lambda_{\chi\chi\mathcal{O}}$ \end{tabular} \right ) + \sum_{\mathcal{O}^-} \lambda^2_{\sigma\epsilon\mathcal{O}} V^{(1)}_{\Delta, \ell} + \lambda^2_{\epsilon\chi\mathcal{O}} V^{(3)}_{\Delta, \ell} + \sum_{\mathcal{O}^+} \lambda^2_{\sigma\chi\mathcal{O}} V^{(2)}_{\Delta, \ell} = 0 \; . \label{app-copy}
\end{equation}
The components of $V^{(1)}$, $V^{(2)}$ and $V^{(3)}$ live in $\mathbb{R}$, while the components of $V^{(0)}$ live in $\mathbb{R}^{3 \times 3}$. To rule out potential solutions, the numerical bootstrap searches for a functional that gives a positive-definite matrix when acting on $V^{(0)}$ and a positive number when acting on the other vectors. Finding such a functional becomes easier when we only demand positivity on specific linear combinations of the vectors above. To this end, we have three options for how to proceed.
\begin{enumerate}
\item If we do not impose \eqref{super-ratio} at all, we use \eqref{app-copy} where the dimensions exchanged by $\sigma \times \chi$ run over all values consistent with unitarity and the presence of three relevant scalar primaries.
\item If we demand the existence of the protected tower discussed in the last section, the last sum in \eqref{app-copy} for odd spin is modified so that it only contains the dimensions of \eqref{change1}.
\item If we use superblocks, we only demand positivity on the linear combinations \eqref{superblocks} instead of their individual components. This means that the last sum in \eqref{app-copy} for even spin is removed altogether and replaced by additional terms in the upper-right and lower-left corners of the matrices in $V^{(0)}$.
\end{enumerate}
In all cases, we impose the basic relations
\begin{eqnarray}
&& \Delta_\sigma + \Delta_\chi = d \nonumber \\
&& \lambda^2_{\sigma\chi\epsilon} = \frac{R_{\sigma\epsilon}}{R_{\chi\epsilon}} \lambda_{\sigma\sigma\epsilon} \lambda_{\chi\chi\epsilon} \label{input}
\end{eqnarray}
which means that the isolated operator $\epsilon$ appears in a superblock. When combined with permutation symmetry, this allows $\sigma$, $\epsilon$ and $\chi$ to be accounted for with a single entry to the first sum of \eqref{app-copy}. Denoting the $m, n$ component of a matrix by $[M]_{mn}$, this entry is
\begin{equation}
\left ( \begin{tabular}{ccc}
$\left [ V^{(0)}_\epsilon \right ]_{11} + V^{(1)}_\sigma$ & $\left [ V^{(0)}_\epsilon \right ]_{12}$ & $\left [ V^{(0)}_\epsilon \right ]_{13} + \frac{1}{2} \frac{R_{\sigma\epsilon}}{R_{\chi\epsilon}} \left ( V^{(1)}_\chi + V^{(2)}_\epsilon + V^{(3)}_\sigma \right )$ \\
$\left [ V^{(0)}_\epsilon \right ]_{21}$ & $\left [ V^{(0)}_\epsilon \right ]_{22}$ & $\left [ V^{(0)}_\epsilon \right ]_{23}$ \\
$\left [ V^{(0)}_\epsilon \right ]_{31} + \frac{1}{2} \frac{R_{\sigma\epsilon}}{R_{\chi\epsilon}} \left ( V^{(1)}_\chi + V^{(2)}_\epsilon + V^{(3)}_\sigma \right )$ & $\left [ V^{(0)}_\epsilon \right ]_{32}$ & $\left [ V^{(0)}_\epsilon \right ]_{33} + V^{(3)}_\chi$
\end{tabular} \right ) \; . \label{single-point}
\end{equation}
The upper-right and lower-left corners account for \eqref{input}, while the upper-left and lower-right corners guarantee $\lambda_{\sigma\sigma\epsilon} = \lambda_{\sigma\epsilon\sigma}$ and $\lambda_{\chi\chi\epsilon} = \lambda_{\epsilon\chi\chi}$ respectively. Treating $\epsilon$ this way leads to interesting bounds on LRIs but it requires all three relevant deformations to be external operators. As we will see shortly, the standard system for the 3D Ising bootstrap --- $\left < \sigma\sigma\sigma\sigma \right >$, $\left < \sigma\sigma\epsilon\epsilon \right >$ and $\left < \epsilon\epsilon\epsilon\epsilon \right >$ --- is not enough.

Scanning over the dimensions of the lightest scalars, our results rely on the \textit{unreasonable effectiveness} of the bootstrap --- the assumption that an interesting theory will lie on the boundary of an excluded region. To carry out the computations, we approximate conformal blocks $G_{\Delta, \ell}^{\Delta_{ij}, \Delta_{kl}}(u, v)$ using the methods of \cite{hr13, hor13, kps13}. While the full details are given in Appendix \ref{sec:appa}, it is useful at this point to mention that these special functions are written as a certain double power series in two variables near the crossing symmetric point $u = v = \frac{1}{4}$. Truncating this expansion requires two cutoffs $(m_{\mathrm{max}}, n_{\mathrm{max}})$. The values chosen in this work are $(3, 5)$, $(5, 7)$ and $(7, 9)$ which correspond to 54, 104 and 170 components respectively.

Since their initial exploration in \cite{kps14}, matrix crossing equations like \eqref{app-copy} have played an increasingly central role in the numerical bootstrap \cite{s15, kpsv15, ll16, b17, no16, kpsv16, ls17a, lms17, lmm18}. They appear whenever there are operators of differing dimension in the four-point functions being analyzed. They have also appeared in the single correlator problem of \cite{bbclp18} which had enough global symmetry for the same representation to be exchanged multiple times.\footnote{Another interesting situation is the long multiplet bootstrap \cite{cls17}. In this case, a mixed system of conformal primaries looks like a single correlator when all parts are combined into superfields. The recent progress \cite{rs18, ahp18} for the superconformal bootstrap in three dimensions appears to be a partial implementation of this idea.} We believe that this is the first time a six-correlator system has been bootstrapped.

\subsection{One correlator}
It is easiest to start with the results that can be obtained from the $\left < \sigma\sigma\sigma\sigma \right >$ correlator alone. In this case, there is no compelling reason to restrict our analysis to three dimensions. Our bound on the spin-2 gap $\Delta_T$, which we plot for 2D and 3D, has been known since the early work in \cite{epprsv12}.
\begin{figure}[h]
\centering
\subfloat[][2D]{\includegraphics[scale=0.4]{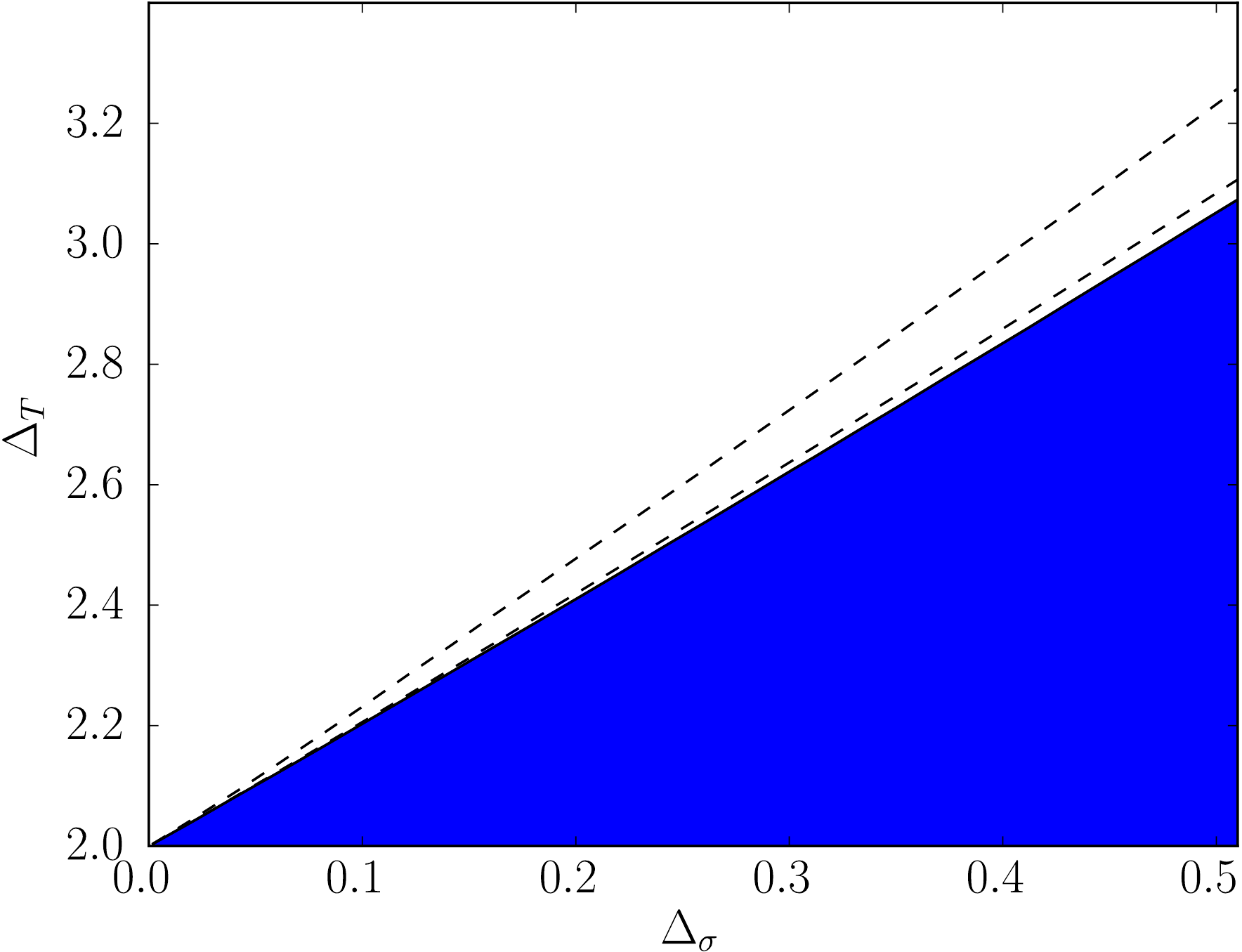}}
\subfloat[][3D]{\includegraphics[scale=0.4]{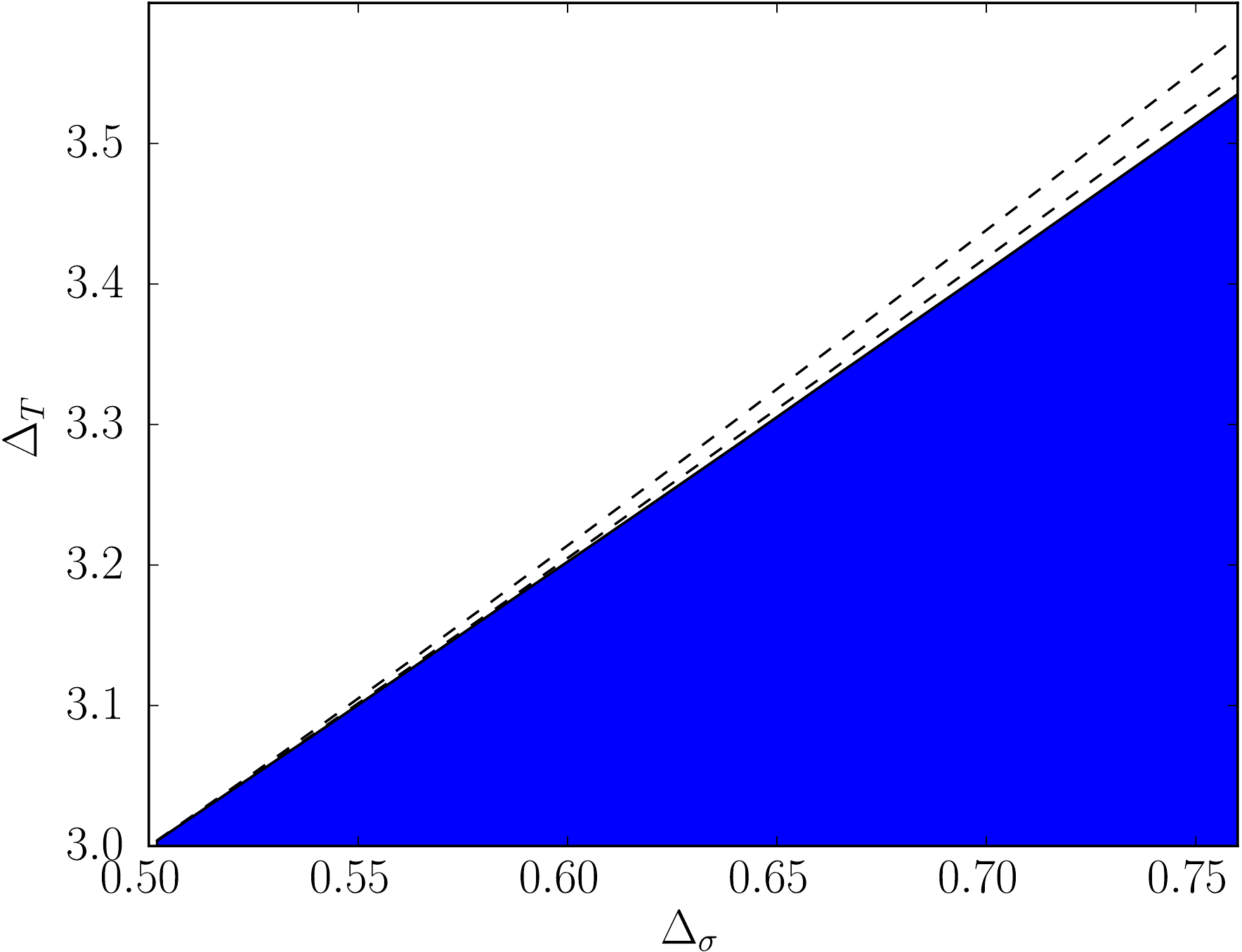}}
\caption{The allowed region for the first spin-2 operator dimension $\Delta_T$ as a function of $\Delta_\sigma$. In both 2D and 3D, the bound appears to be converging to $\Delta_T = 2\Delta_\sigma + 2$. The blue region was obtained with $(m_{\mathrm{max}}, n_{\mathrm{max}}) = (7, 9)$ while $(5, 7)$ and $(3, 5)$ are shown for comparison.}
\label{fig:dimt}
\end{figure}

From Figure \ref{fig:dimt}, it appears that the $\Delta_T$ bound is saturated by generalized free field theory. This gives us an idea of how the allowed region in $(\Delta_\sigma, \Delta_\epsilon)$ space must behave. Not only must it become more restrictive as $\Delta_T$ is increased --- its boundary must move from left to right at a known rate.

\begin{figure}[h]
\centering
\subfloat[][2D]{\includegraphics[scale=0.42]{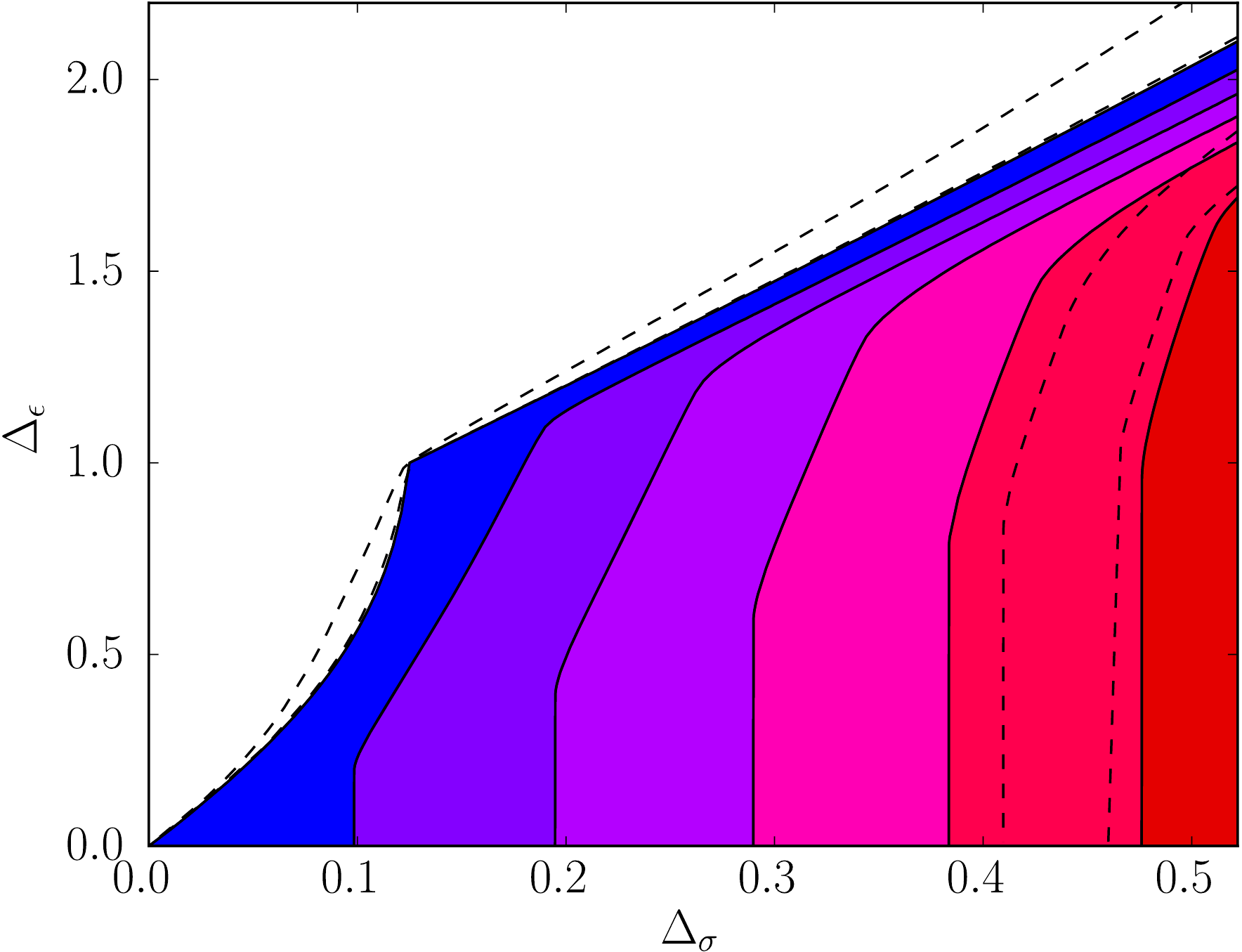}}
\subfloat[][3D]{\includegraphics[scale=0.42]{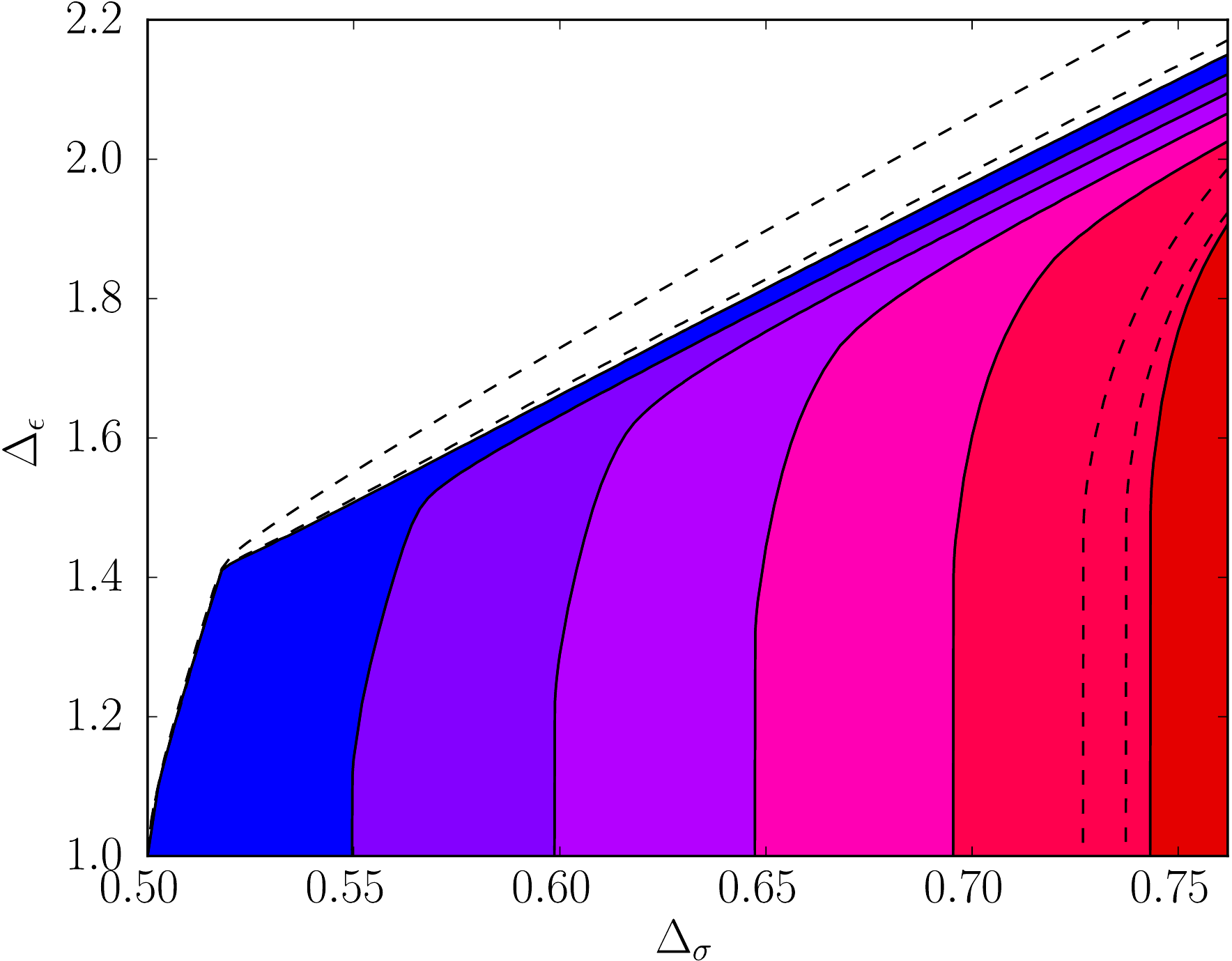}}
\caption{The upper bound on $\Delta_\epsilon$ as a function of $\Delta_\sigma$. Our spin-2 constraint goes from $\Delta_T \geq d$ (blue) to $\Delta_T \geq \frac{d}{2} + 2$ (red) in evenly spaced steps. Again, our main plots have $(m_{\mathrm{max}}, n_{\mathrm{max}}) = (7, 9)$ with dotted lines for $(5, 7)$ and $(3, 5)$. As expected, the convergence of the red region is slower than that of the blue region.}
\label{fig:1corr}
\end{figure}

It is straightforward to derive an upper bound of this type. In Figure \ref{fig:1corr}, we have done this for six different values of the spin-2 gap. In 2D, the minimum $\Delta_T$ values we sample are $\{2, 2.2, 2.4, 2.6, 2.8, 3\}$, while in 3D they are $\{3, 3.1, 3.2, 3.3, 3.4, 3.5\}$. If the previously observed saturation is correct, the edges of these plots must continue moving left as our computational power is incresaed. For instance, we expect a $5\%$ change for the 2D red region and a $3\%$ change for the 3D red region.
\begin{figure}[h]
\centering
\subfloat[][Spin-0]{\includegraphics[scale=0.42]{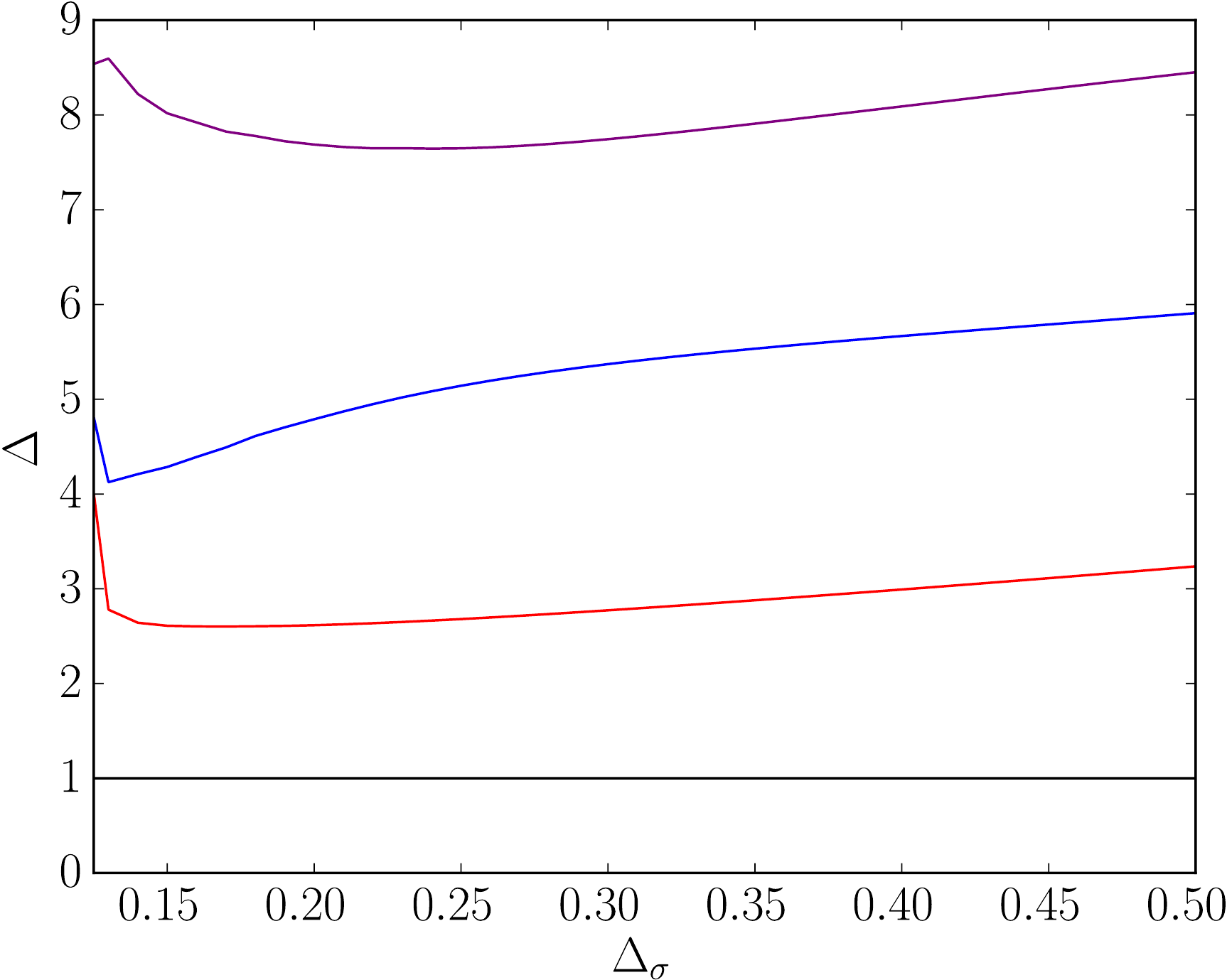}}
\subfloat[][Spin-2]{\includegraphics[scale=0.42]{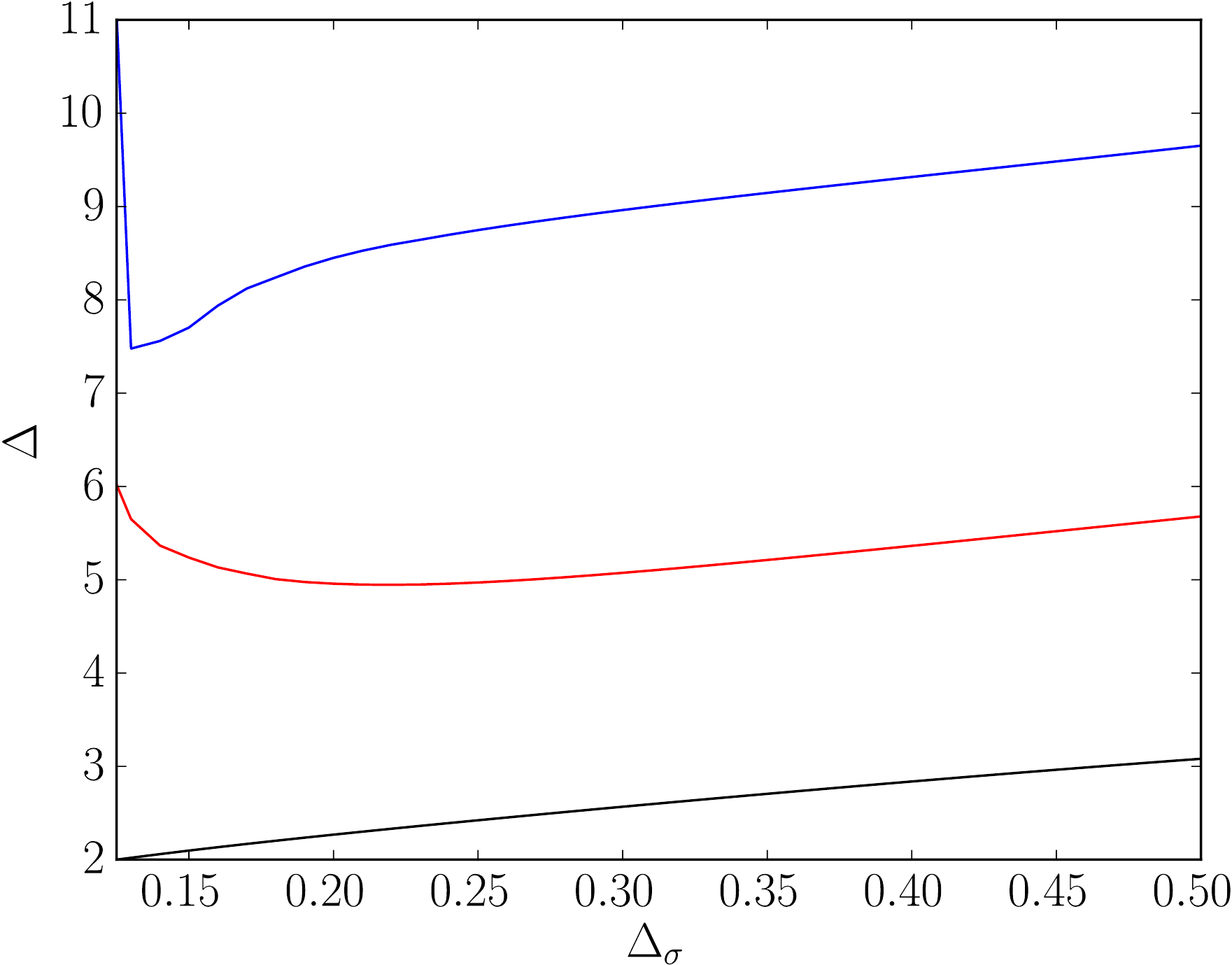}}
\caption{A few scaling dimensions in the extremal spectrum having maximal $\lambda^2_{\sigma\sigma\epsilon}$ with $\Delta_\epsilon = 1$. These are zeros of the functional that is found during the maximization procedure. For all of the spins that we have tested, it is plausible that these could approximate primary operators in the LRI.}
\label{fig:extremal}
\end{figure}

Once we disallow $\Delta_T$ at the unitarity boumd, the kink corresponding to the SRI quickly disappears. The point $(\Delta_\sigma, \Delta_\epsilon) = \left ( \frac{d}{4}, \frac{d}{2} \right )$, marking the first LRI to be described by mean-field theory, does not display any feature. We may therefore conclude that a single correlator gives very little information about the spectrum of non-trivial long-range Ising models.

Before proceeding to our multi-correlator results in three dimensions, we should comment on the fact that some theories can saturate numerical bootstrap bounds even when there is no kink. For a judiciously chosen quantity, there is some evidence that this is the case for the LRI in two dimensions.\footnote{This idea is due to Sheer El-Showk.} Instead of bounding a gap, one can maximize the OPE coefficient of an operator in the spectrum. From Monte Carlo data \cite{apr14}, it is clear that in every 2D long-range Ising model, one such operator has a dimension close to $1$. In some sense, this is explained by the perturbative calculation \eqref{dim-eps2} around $s = s_*$, since the leading order anomalous dimension of $\epsilon$ happens to vanish. By setting $\Delta_\epsilon = 1$ and maximizing $\lambda_{\sigma\sigma\epsilon}^2$, we have extracted the low-lying spectrum using the extremal functional method of \cite{ps11, ep13, ep16}. At $\Delta_\sigma = \frac{1}{8}$, this is guaranteed to agree with the spectrum of the SRI. However, Figure \ref{fig:extremal} also shows very interesting behaviour at $\Delta_\sigma = \frac{1}{2}$ as we now explain.\footnote{For this step, we have used the \texttt{spectrum.py} script of \cite{s17} with parameters $(m_{\mathrm{max}}, n_{\mathrm{max}}) = (5, 7)$. This appears to be close to the threshold of where the script can produce reasonable results. Setting $m_{\mathrm{max}} = 4$ for example, we find approximately the same scaling dimensions but OPE coefficients that are off by ten orders of magnitude.}

Here, the leading spin-0 and spin-2 dimensions are both close to $3$. We can see that this is exactly what happens in a generalized free field theory if we write the operators as $\sigma \partial^2 \sigma$ and $\sigma \partial_\mu \partial_\nu \sigma$ respectively. The dimensions of the next double-twist operators, found by inserting extra powers of $\partial^2$, appear somewhat too high but this could easily be an effect of the numerics. This makes it tempting to conjecture that given a long-range Ising model with dimensions $(\Delta_\sigma, \Delta_\epsilon)$, all other crossing symmetric four-point functions $\left < \sigma\sigma\sigma\sigma \right >$ have a smaller value of $\lambda_{\sigma\sigma\epsilon}^2$. This approach to studying the LRI is ultimately perturbative since it requires the dimension of $\epsilon$ as input. Nevertheless, it could be useful for reducing the number of Feynman diagrams one encounters. Instead of computing separate diagrams for each anomalous dimension, the conjecture would enable us to compute only diagrams for $\gamma_\epsilon$ and then feed these into the bootstrap machinery to learn about other observables.
\begin{table}[h]
\centering
\begin{tabular}{c|c}
Operator & Dimension \\
\hline
$\epsilon$ & $1$ \\
$\sigma\chi$ & $2$ \\
$\chi^2$ & $\frac{15}{4}$ \\ 
$\epsilon^\prime$ & $4$ \\
$\epsilon \chi^2$ & $\frac{19}{4}$ \\ 
$\sigma \chi^3$ & $\frac{23}{4}$ \\ 
$\chi \partial^2 \chi$ & $\frac{23}{4}$ \\ 
$\sigma^\prime \chi$ & $6$
\end{tabular}
\quad
\begin{tabular}{c|c}
Operator & Dimension \\
\hline
$\epsilon \chi \partial^2 \chi$ & $\frac{27}{4}$ \\ 
$\chi^4$ & $\frac{15}{2}$ \\ 
$\epsilon^\prime \chi^2$ & $\frac{31}{4}$ \\ 
$\sigma \chi^2 \partial^2 \chi$ & $\frac{31}{4}$ \\ 
$\chi \partial^4 \chi$ & $\frac{31}{4}$ \\ 
$\epsilon^{\prime\prime}$ & $8$ \\
$\epsilon \chi^4$ & $\frac{17}{2}$ \\ 
$\epsilon \chi \partial^4 \chi$ & $\frac{35}{4}$ 
\end{tabular}
\caption{Some operators having $\Delta < 9$ in the CFT obtained by coupling the 2D SRI to a generalized free field of dimension $\frac{15}{8}$. We only show the ones where both parts are scalars. Anything involving $\chi$ decouples from $\sigma \times \sigma$ at $s = s_*$ where the quoted dimensions hold exactly. At slightly smaller values of $s$ however, these operators acquire a nonzero OPE coefficient as long as they are even with respect to the diagonal $\mathbb{Z}_2$ from the two theories.}
\label{missing-ops}
\end{table}

The sparseness of the spectrum in Figure \ref{fig:extremal} hints at another significant limitation. By perturbing around $s = s_*$ or $s = \frac{d}{2}$, it becomes clear that several additional families of operators enter the $\sigma \times \sigma$ OPE in a generic LRI. In particular, the number of scalars having $\Delta < 9$ should be much more than 4. Table \ref{missing-ops} shows 16 such operators that can be constructed with the deformation of \cite{brrz17a, brrz17b}.

The tendency for the extremal functional method to miss several operators was discussed in \cite{s17}, which noticed that the numerical spectrum is dominated by double-twist families. These happen to be the families required to match the crossed-channel singularity produced by a unique minimal-twist operator in the analytic bootstrap of \cite{fkps13, kz13}. A loose conjecture arising from this is that in any crossing equation with a twist gap, several multi-twist operators with significant OPE coefficients will nevertheless provide a negligible contribution in the numerical boostrap. This is supported, for instance, by the test of the extremal functional method in \cite{b18}, showing essential differences between the 2D and 3D Ising models. So far, the most reliable numerical bootstrap spectra all come from special cases involving a higher spin symmetry. It is worth mentioning that the analytic bootstrap has recently been extended to handle these cases as well in \cite{a17, c17}. What this means for the present case is that we only get a clear picture of the low-lying operators at $\Delta_\sigma = \frac{1}{8}$ \textit{i.e.} the 2D Ising spectrum. As soon as we raise $\Delta_\sigma$, the theory maximizing $\lambda_{\sigma\sigma\epsilon}^2$ becomes nonlocal. Even though the nonlocality is small, we never see operators involving $\chi$ because their contributions in the $\sigma \times \sigma$ OPE are small as well.

\subsection{Three correlators}
\begin{figure}[h]
\centering
\includegraphics[scale=0.7]{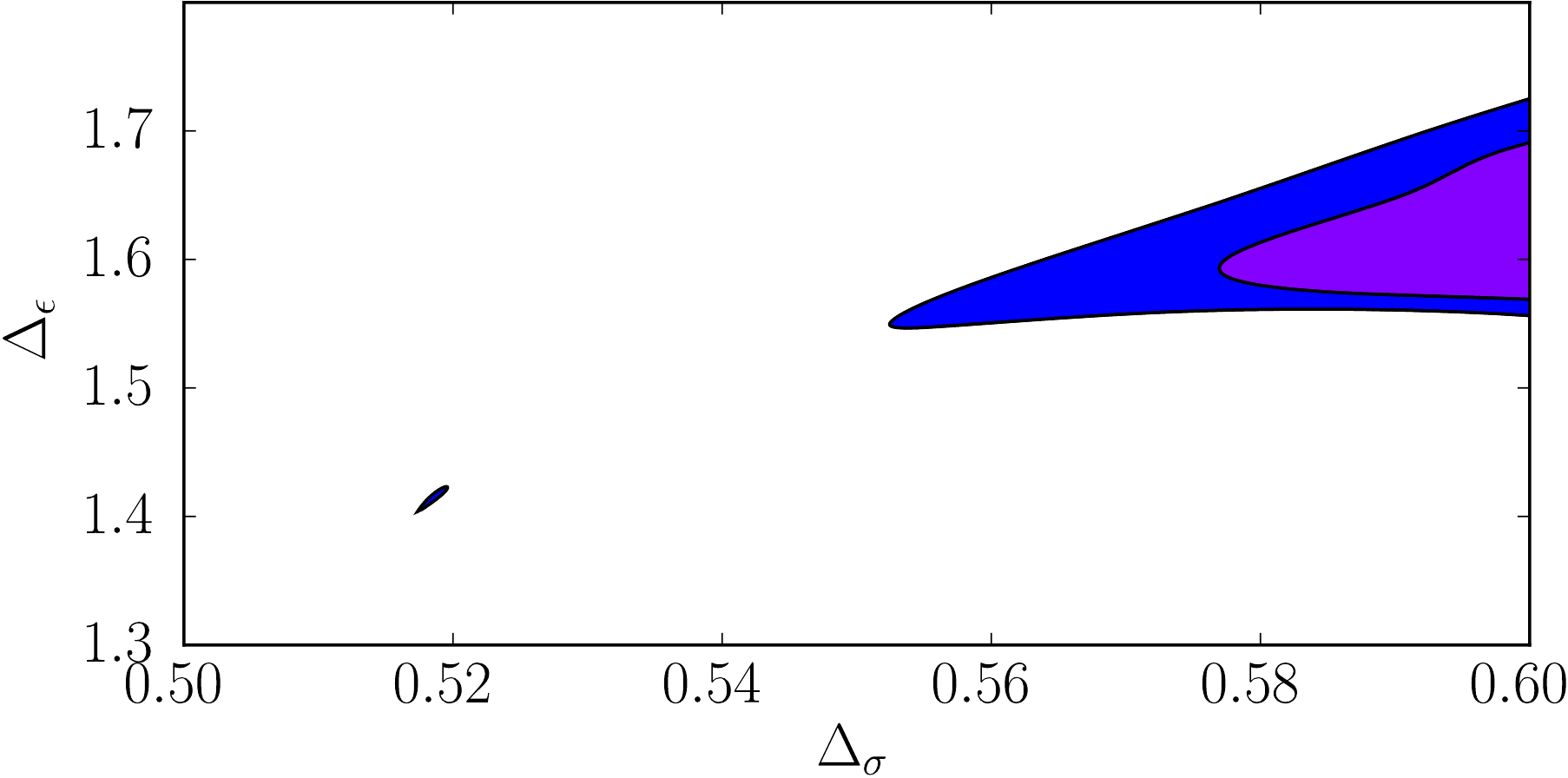}
\caption{Constraints on the space of CFTs with one relevant primary operator of each parity. The allowed region for $\Delta_T \geq 3$ is blue, while the one for $\Delta_T \geq 3.1$ is purple. In the former case, an island around the 3D Ising model is separated from the rest of the region. This excludes many long-range Ising models which require $\chi$ to be present.}
\label{fig:local}
\end{figure}
\begin{figure}[h]
\centering
\includegraphics[scale=0.7]{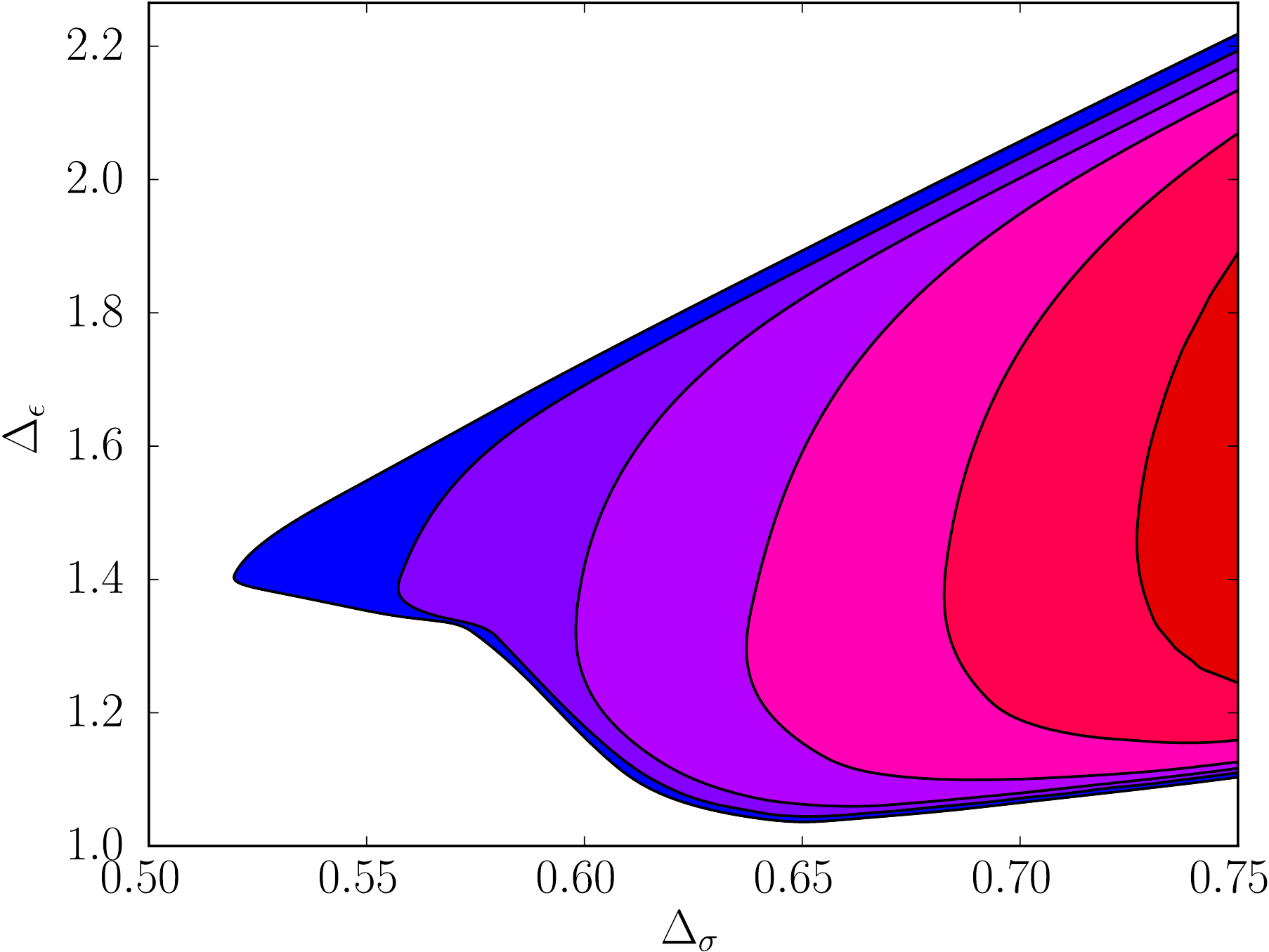}
\caption{A multi-correlator version of Figure \ref{fig:1corr}, computed with $(m_{\mathrm{max}}, n_{\mathrm{max}}) = (3, 5)$. The upper bounds are similar to the ones plotted before but the lower bounds are new. We have plotted them using a simple bisection while also testing interior points to ensure that there are no holes. Again, blue signifies $\Delta_T \geq 3$ and red signifies $\Delta_T \geq 3.5$.}
\label{fig:3corr}
\end{figure}
\begin{figure}[h]
\centering
\includegraphics[scale=0.7]{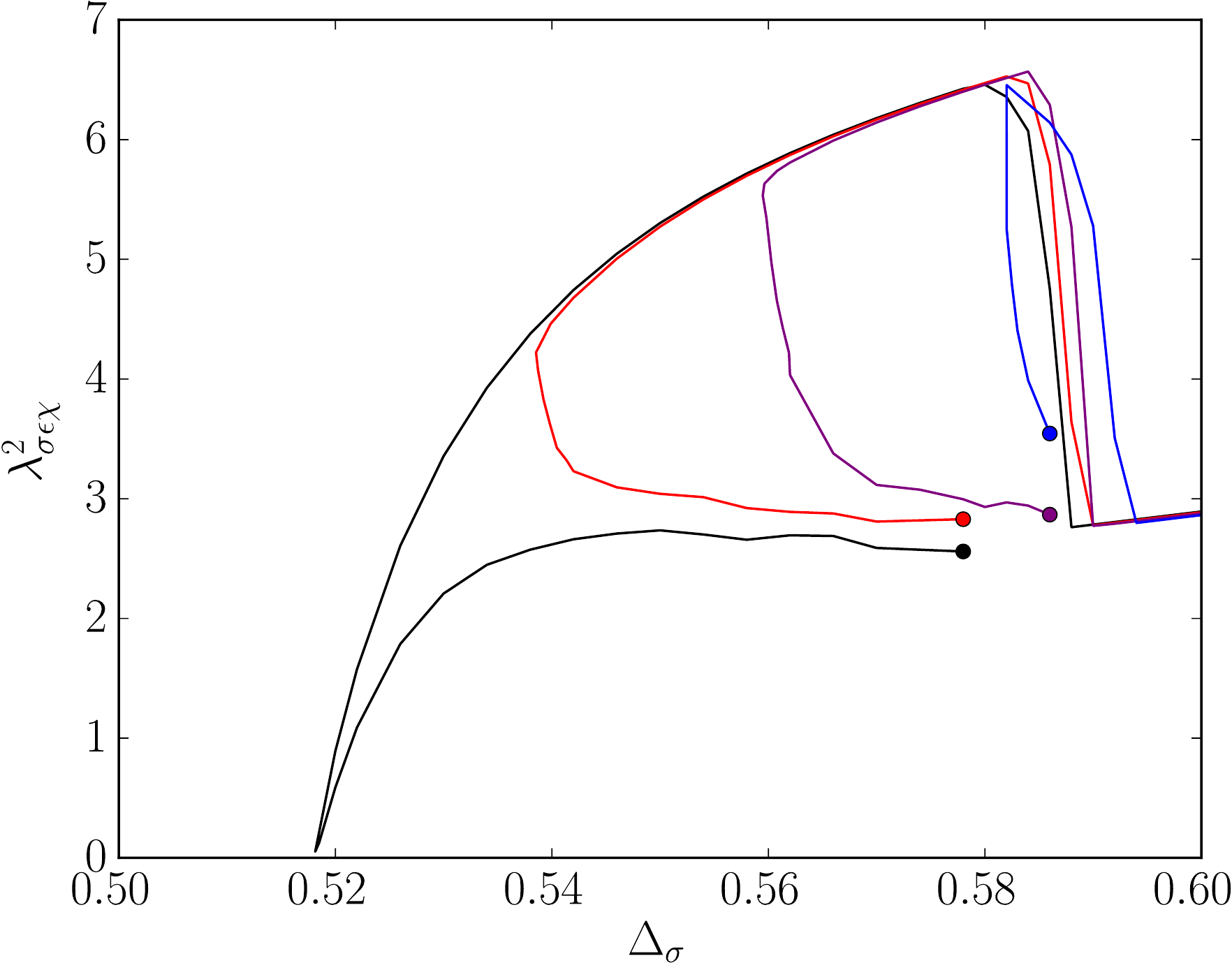}
\caption{The bound on $\lambda_{\sigma\epsilon\chi}^2$ as a function of $\Delta_\sigma$ computed for $(m_{\mathrm{max}}, n_{\mathrm{max}}) = (5, 7)$. The minimum $\Delta_T$ is $3$ for the black line, $3.05$ for the red line, $3.1$ for the purple line and $3.15$ for the blue line. The OPE coefficient is maximized along the upper and lower branches of Figure \ref{fig:3corr}. However, we have chosen not to go all the way to $\Delta_\sigma = 0.6$ along the upper branch. Doing so would yield several intersecting lines that reduce visual clarity.}
\label{fig:decoupling}
\end{figure}
In order to improve upon our single correlator results, the next logical step is to bootstrap the correlators $\left <\sigma\sigma\sigma\sigma \right >$, $\left < \sigma\sigma\epsilon\epsilon \right >$ and $\left < \epsilon\epsilon\epsilon\epsilon \right >$. For this system, it makes a difference whether there are two relevant primary operators or three. Even if we did not know about the shadow relation \eqref{non-renorm2}, we would be able to infer the existence of a third relevant primary from the well known island of \cite{kps14}. The assumptions that lead to an island are incompatible with the LRI because there must be a continuous line of fixed points that lead away from the SRI.

Imposing the existence of three relevant primaries, as we should, we find a reassuring exclusion plot in which all regions are connected. Figure \ref{fig:3corr} shows how they change as a function of $\Delta_T \in \{ 3, 3.1, 3.2, 3.3, 3.4, 3.5 \}$. Because $3 - \Delta_\sigma$ is different from $3\Delta_\sigma$, many of the generalized free theory solutions from Figure \ref{fig:1corr} are excluded this time.

The boundary of each region has an upper branch and a lower branch. For the intermediate values $\Delta_T \in \{ 3.1, 3.2, 3.3, 3.4 \}$, we do not observe any evidence that either branch contains a point corresponding to an LRI. If intermediate LRI models do saturate one of the branches, existing estimates for the critical exponents suggest that this should be the upper one. Lower branches for these values of $\Delta_T$ all have $\Delta_\epsilon < 1.4$. While there is no candidate LRI kink in Figure \ref{fig:3corr}, there is a ``concave kink'' for some of the lower branches at $\Delta_\sigma \approx 0.58$. It appears to be a coincidence that the leftmost edge of the $\Delta_T \geq 3.1$ region of Figure \ref{fig:local} is also near this value of $\Delta_\sigma$. If any bound in Figure \ref{fig:3corr} were to intersect the region where CFTs can exist without $\chi$, a vanishing $\lambda_{\sigma\epsilon\chi}^2$ would signal the presence of a kink. Instead, we have found that this OPE coefficient decreases slightly at the special point without going to zero. In Figure \ref{fig:decoupling}, we maximize $\lambda_{\sigma\epsilon\chi}^2$ for $\Delta_T \in \{ 3, 3.05, 3.1, 3.15 \}$.\footnote{Maximizing an OPE coefficient helps to reduce any error that might have been introduced by our $10^{-4}$ bisection threshold. Once the boundary is found with sufficient precision, the spectrum is already uniquely fixed and extremization procedures are superfluous.} The fact that $\chi$ decouples at a single point in the local case supports the proposal in \cite{brrz17a}. It also agrees with the expectation that there is only one irreducible CFT with the same critical exponents as the Ising model.

\subsection{Six correlators}
In order to gain non-perturbative information about the LRI critical exponents, we will need to examine the minimal system of four-point functions that allows access to \eqref{input}. This consists of $\left < \sigma\sigma\epsilon\epsilon \right >$, $\left < \sigma\sigma\chi\chi \right >$, $\left < \epsilon\epsilon\chi\chi \right >$ and the three identical correlators. This system yields a much more restrictive region than Figure \ref{fig:3corr} and it will turn out to have interesting features. In order to check that they are in the right place, we have summarized our perturbative data about the LRI in Table \ref{3d-data}. The last row resums the expansions around $s = \frac{d}{2}$ and $s = s_*$ using the $[3, 3]$ Pad\'e approximant. What this means is that we start with the ansatz
\begin{equation}
\Delta_{\mathcal{O}}(s) = \frac{a_0 + a_1 s + a_2 s^2}{1 + b_1 s + b_2 s^2} \; , \label{pade-ansatz}
\end{equation}
and fix the coefficients by demanding that \eqref{pade-ansatz} have the correct Taylor expansion around the two solvable points.
\begin{table}[h]
\centering
\begin{tabular}{c|c|c}
& $\Delta_\epsilon$ & $\Delta_T$ \\
\hline
$\varepsilon$-expansion & $\frac{3}{2} - \frac{1}{6} \varepsilon + 0.18122 \varepsilon^2$ & $\frac{7}{2} - \frac{1}{2} \varepsilon - 0.05926 \varepsilon^2$ \\
$\delta$-expansion & $1.41263 + 0.269 \delta$ & $3 + 2.333 \delta$ \\
$[3, 3]$-Pad\'e & $\frac{1.3759 - 1.7116s + 0.4013s^2}{1 - 1.2086s + 0.2758s^2}$ & $\frac{4.7026 - 4.3183s + 0.8191s^2}{1 - 0.7812s + 0.0850s^2}$
\end{tabular}
\caption{Restating our perturbative results for unprotected operators in the LRI. These expressions are specializations of \eqref{dim-eps}, \eqref{dim-t}, \eqref{dim-eps2} and \eqref{dim-t2} to $d = 3$. To interpolate between the two expansions, we have calculated the symmetric Pad\'e approximant.}
\label{3d-data}
\end{table}

\begin{figure}[h]
\centering
\includegraphics[scale=0.7]{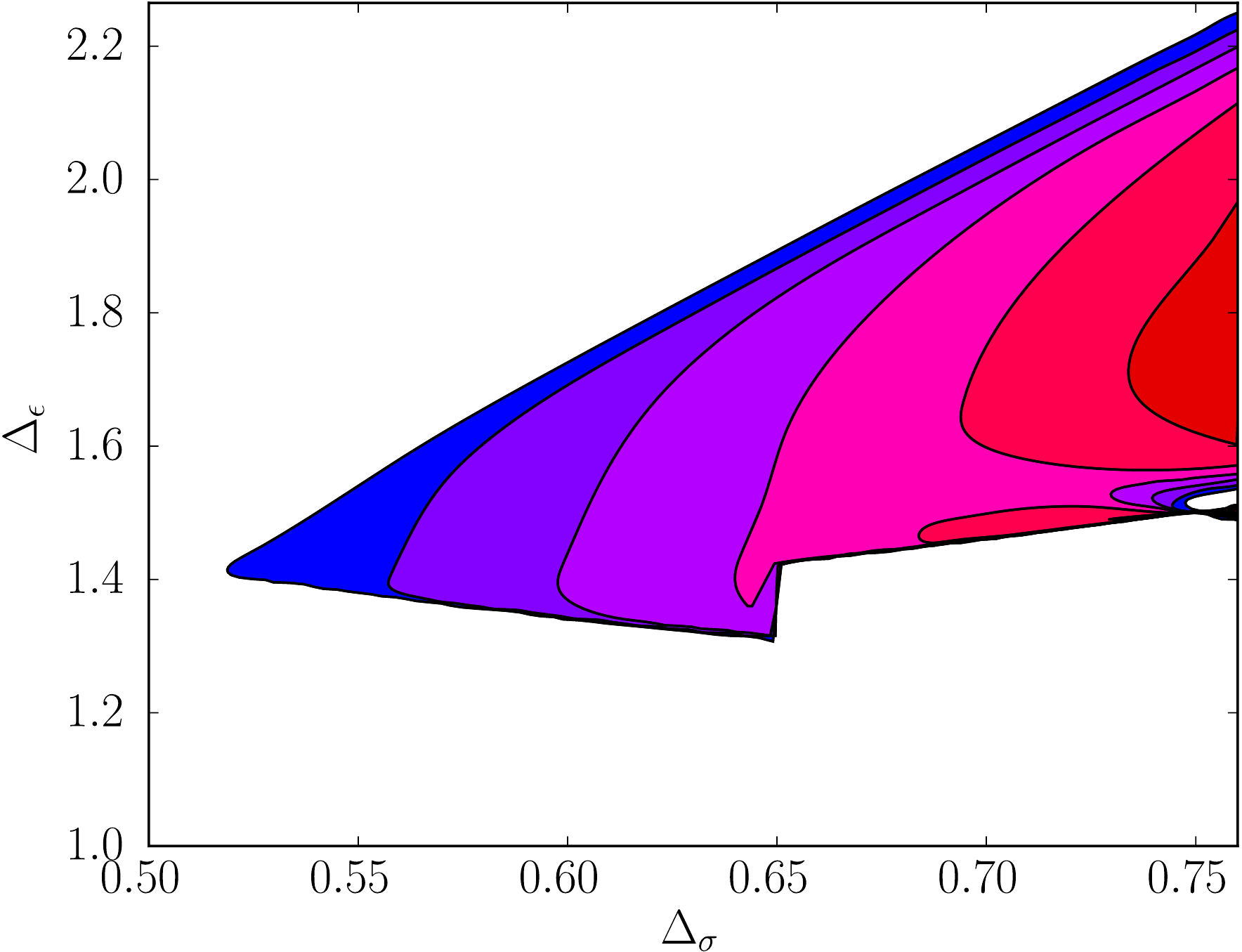}
\caption{The allowed region in $\left ( \Delta_\sigma, \Delta_\epsilon \right )$ space found by imposing crossing symmetry and unitarity on the six correlator system that includes $\sigma$, $\epsilon$ and $\chi$. As in Figure \ref{fig:3corr}, the most permissive region (the blue one) allows the first $\mathbb{Z}_2$-even spin-2 operator dimension $\Delta_T$ to be as low as $3$, while the most restrictive region (the red one) forces it to be at least $3.5$. The other regions have $\Delta_T \in \{ 3.1, 3.2, 3.3, 3.4 \}$. Derivative orders are $(m_{\mathrm{max}}, n_{\mathrm{max}}) = (3, 5)$. These regions account for protected double-twist operators in $\sigma \times \chi$ and use ordinary conformal blocks which means that the $\left < \sigma\sigma\chi\chi \right >$ crossing equations are those of \eqref{change1}.}
\label{fig:6corr}
\end{figure}
Excluding points for $\Delta_T \in \{ 3, 3.1, 3.2, 3.3, 3.4, 3.5 \}$ again reveals the boundaries in Figure \ref{fig:6corr}. The lower branches are much more restrictive than those in Figure \ref{fig:3corr} even though we have not made use of the superblocks yet. As an example, the plot for $\Delta_T = 3$ already appears to single out the onset of mean-field theory --- the blue region reaches a very narrow throat at $\left ( \Delta_\sigma, \Delta_\epsilon \right ) = \left ( \frac{3}{4}, \frac{3}{2} \right )$. The lower branch for this plot also experiences a jump at $\Delta_\sigma \approx 0.65$.\footnote{The jump here would be less pronounced if we did not assume protected operators at dimensions given by \eqref{super-ratio-poles}. To the left of $\Delta_\sigma \approx 0.65$, it appears to make no difference whether we impose the existence of this tower or not. We believe that most of the constraints here come from the OPE coefficient relations for $\epsilon$ that are captured in \eqref{single-point}.} These features persist for higher values of $\Delta_T$ as well until the allowed region splits into two lobes. The bottom lobe of the $\Delta_T = 3.5$ plot stays very narrow to the left of the throat and ends at $\Delta_\sigma \approx 0.73$. This leftmost edge continues to recede as the number of derivatives is increased.

The regions shown here start to look more promising after we increase the number of derivatives. The $\Delta_T = 3.3$ region, for instance, moves to the right of the jump and develops two lobes that are connected by a narrow bridge. This makes it possible to plot a comparison between the bottom lobes and the results of Table \ref{3d-data} for $\Delta_T > 3.25$. Instead of performing this check \textit{ceteris paribus}, we have removed the assumption that the $[\sigma\chi]_{n, \ell}$ operators are protected. The main conjecture of this work can be tested after the fact by computing an extremal spectrum at several points using the script in \cite{s17}.
\begin{figure}[h]
\centering
\includegraphics[scale=0.7]{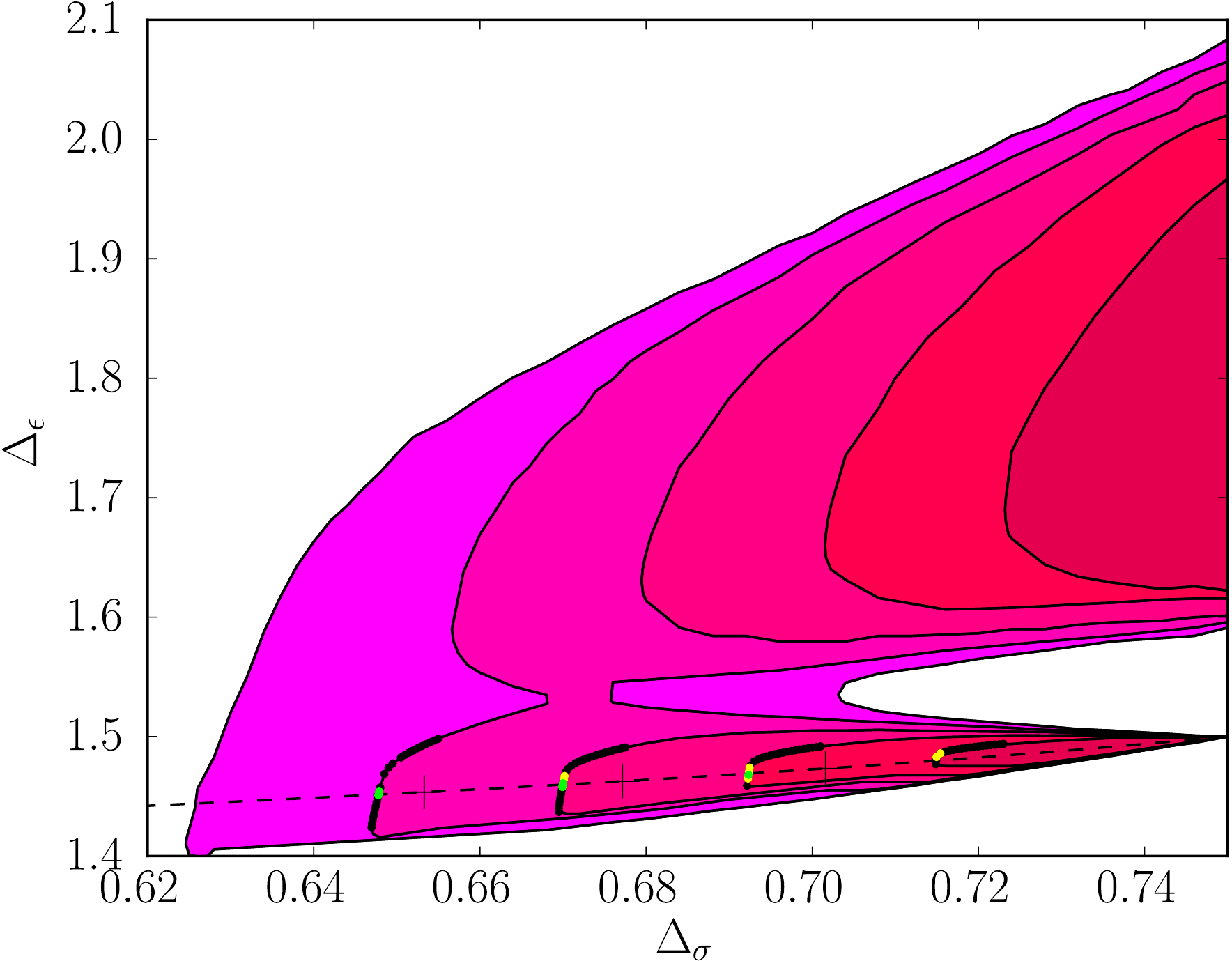}
\caption{Five allowed regions whose spin-2 restrictions increment from $\Delta_T \geq 3.25$ on the left to $\Delta_T \geq 3.45$ on the right. They were found using ordinary conformal blocks with $(m_{\mathrm{max}}, n_{\mathrm{max}}) = (5, 7)$. Since no protected operators were assumed, the relevant crossing equation for $\left < \sigma\sigma\chi\chi \right >$ is \eqref{change0}. Green points contain the operator $(\Delta, \ell) = (\Delta_T, 2)$ in all $\mathbb{Z}_2$-even OPEs instead of just one of them. Green and yellow points contain a $\mathbb{Z}_2$-even vector of dimension close to $6$. All points that we have found to have neither property are marked in black. The perturbative dotted line shows $\Delta_\epsilon$ as a function of $\Delta_\sigma$ according to the Pad\'e approximant in Table \ref{3d-data}. Points on this line that are predicted to have $\Delta_T \in \{3.3, 3.35, 3.4\}$ are denoted by crosses.}
\label{fig:zoom}
\end{figure}

Allowing a continuum of odd-spin operators in the $\sigma \times \chi$ OPE we have plotted the allowed regions for $\Delta_T \in \{ 3.25, 3.3, 3.35, 3.4, 3.45 \}$ in Figure \ref{fig:zoom}. Points on the edge, where we have extracted the spectrum, have been highlighted if they exhibit one of the following interesting properties.
\begin{enumerate}
\item
A point is yellow if it contains a vector suitably close to $[\sigma\chi]_{1,1}$. Our threshold is that its dimension must be within $5\%$ of $6$.
\item
A point is green if it additionally contains a symmetric tensor in $\sigma \times \chi$ whose dimension is within $5\%$ of $\Delta_T$. Note that this is always true for the OPEs with Bose symmetry.
\end{enumerate}
\begin{figure}[h]
\centering
\subfloat[][$\Delta_T = 3.3$]{\includegraphics[scale=0.4]{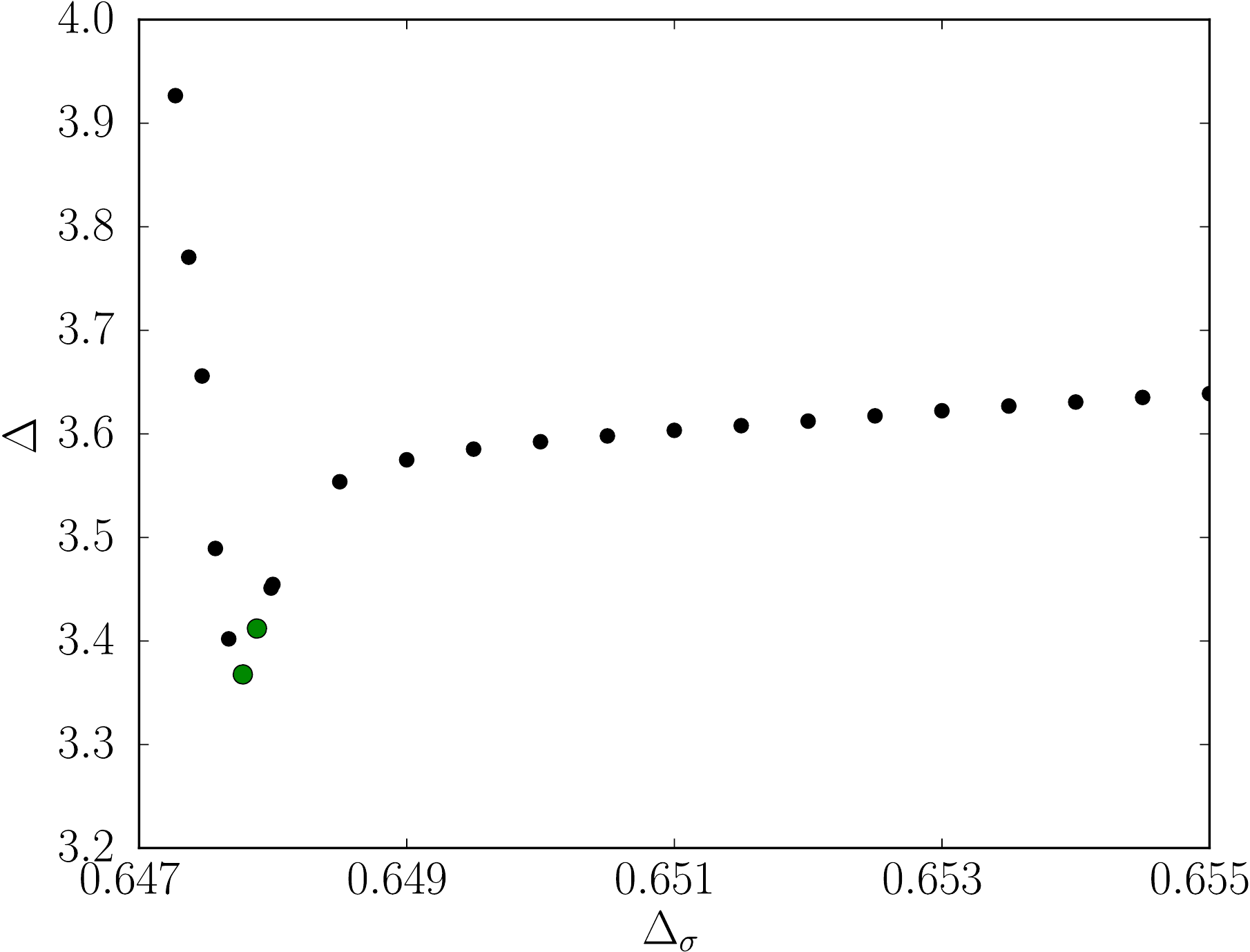}}
\subfloat[][$\Delta_T = 3.35$]{\includegraphics[scale=0.4]{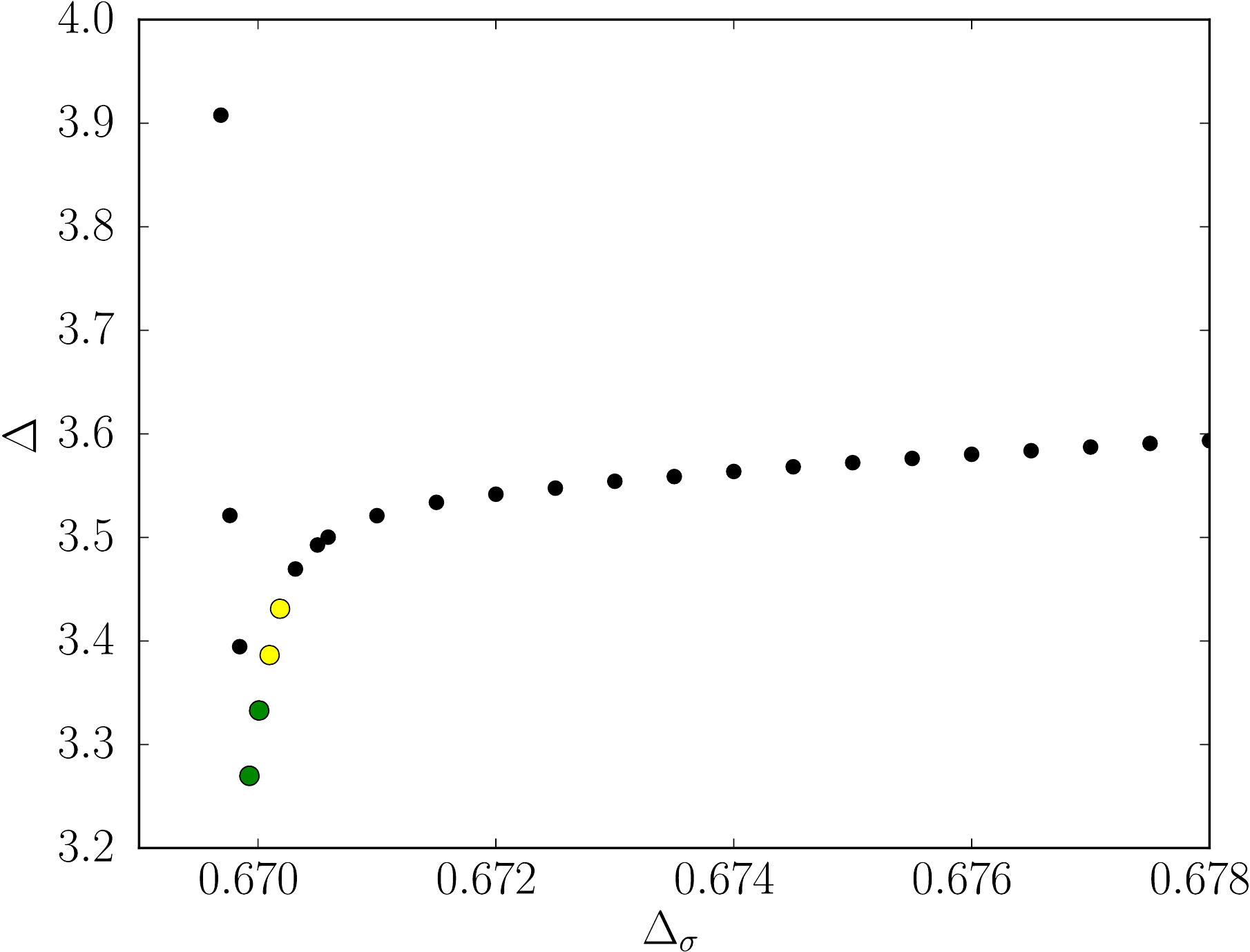}} \\
\subfloat[][$\Delta_T = 3.4$]{\includegraphics[scale=0.4]{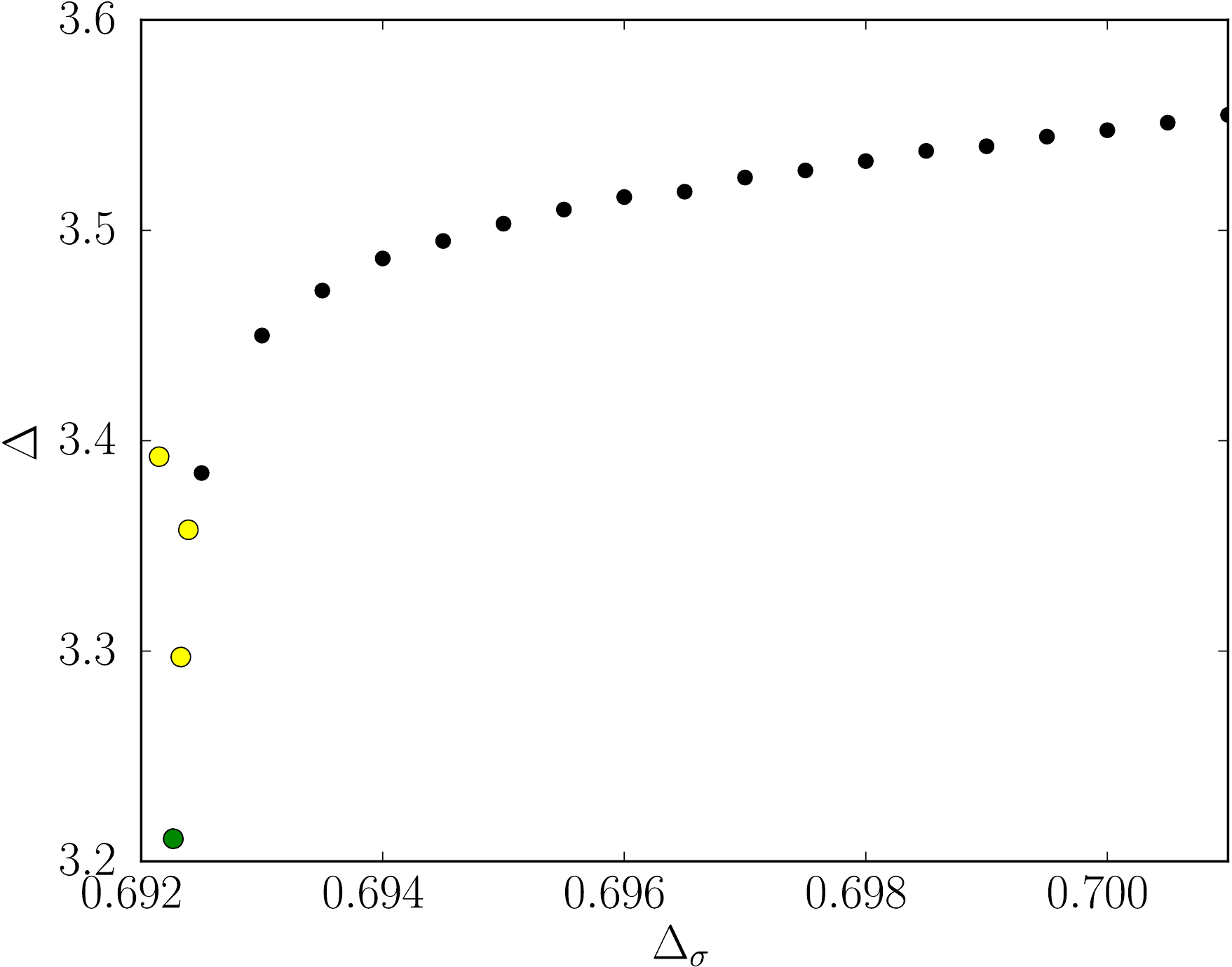}}
\subfloat[][$\Delta_T = 3.45$]{\includegraphics[scale=0.4]{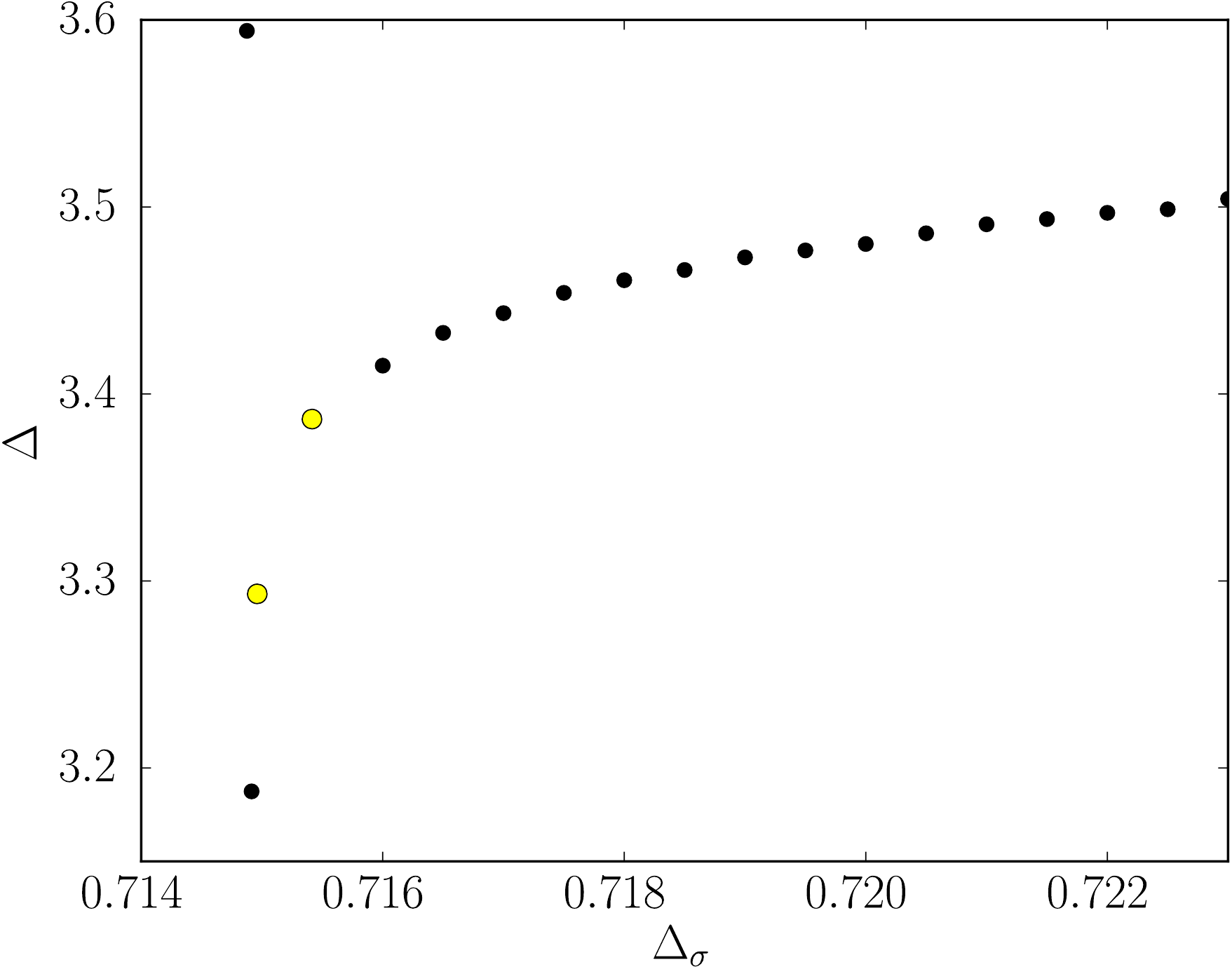}}
\caption{The dimension of the first irrelevant scalar in each of the four spectra extracted at the points shown in Figure \ref{fig:zoom}. These have been taken from the $\sigma \times \sigma$, $\epsilon \times \epsilon$ and $\chi \times \chi$ OPEs as the scalars in $\sigma \times \chi$ do not show an interesting feature. Points that are yellow and green respectively satisfy property 1 and property 2 as defined in the text. These points have also been made larger.}
\label{fig:scalar3}
\end{figure}

Points in the first set are likely to survive when we impose the non-renormalization of the double-twist tower. Points in the second set are likely to survive when we use superblocks or the nine correlator system described in Appendix \ref{sec:appa}.\footnote{Checking what survives by producing another exclusion plot is not always instructive. In many cases, the change in a given bound is not visible to the naked eye.} Further significance to these points can be seen by plotting the dimension of the first irrelevant scalar. In all long-range Ising models, we expect a value reasonably close to $3$ since $\sigma\chi$ is marginally irrelevant at $s = s_*$ and $\phi^4$ is marginally irrelevant at $s = \frac{d}{2}$. Figure \ref{fig:scalar3} shows that this is predominantly achieved at the green and yellow points which cluster around the local minimum. Several other exercises along these lines are possible, \textit{e.g.} checking that the $\mathbb{Z}_2$-odd OPEs $\sigma \times \epsilon$ and $\epsilon \times \chi$ have low-lying operators in common as well.\footnote{The validity of this has nothing to do with the LRI specifically. In any bootstrap problem that includes several external scalars but not all of their mixed correlators, there are going to be OPEs that are not constrained to exchange the same operators despite being identical in terms of representation theory. In such a problem, the possibility of having two disjoint $\mathbb{Z}_2$-even OPEs, for instance, is generic but completely unphysical.} One could also imagine a comparison involving OPE coefficients, in order to see that versions of \eqref{ratio-cancelled} hold with multiple spinning operators. In practice, we have found this difficult as some of the gamma functions are highly sensitive to small errors in the exchanged dimension.

Looking at the $[3, 3]$ Pad\'e approximant line in Figure \ref{fig:zoom}, we see that it comes remarkably close to the green and yellow points. Even after a lobe becomes blunt enough that it can no longer be considered a kink, it is apparently worthwhile to locate desirable features in the extremal spectrum. This approach, advocated in \cite{ep16}, could perhaps reveal useful information about the smooth $\Delta_T < 3.25$ boundaries in Figure \ref{fig:6corr}. The crosses in Figure \ref{fig:zoom}, being offset from the green and yellow points, simply reflect the fact that convergence is noticeably slower near the $\left ( \Delta_\sigma, \Delta_\epsilon \right ) = \left ( \frac{3}{4}, \frac{3}{2} \right )$ mean-field theory. They would appear very close to the boundaries if we continued to plot them down to $\Delta_T = 3$.

\begin{figure}[h]
\centering
\includegraphics[scale=0.7]{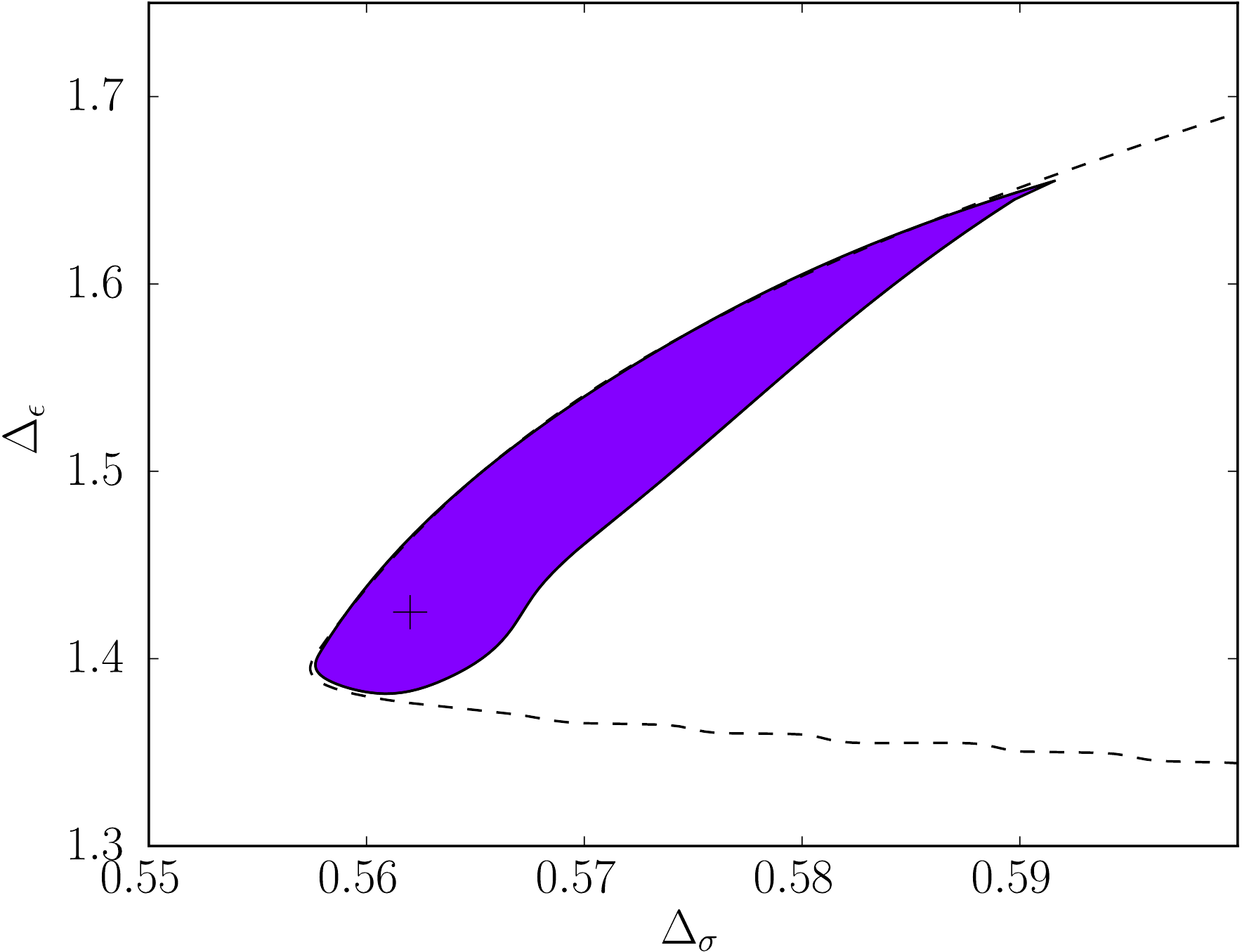}
\caption{The island for the $\Delta_T = 3.1$ model computed with $(m_{\mathrm{max}}, n_{\mathrm{max}}) = (3, 5)$. Unlike Figure \ref{fig:6corr}, which contains a much larger purple region for $\Delta_T = 3.1$, this uses the full content of the shadow relation captured in the crossing equation \eqref{change2} --- \textit{i.e.} it was obtained by demanding crossing symmetry and unitarity for the ansatz built out of the superblocks \eqref{superblocks}. An island only forms because of the superblocks and the fact that we are imposing a spin-2 gap above $\Delta_T$. In this case, the gap is $\Delta_{T^\prime} \geq 4.5$. The old $\Delta_T = 3.1$ region that would be produced from a bootstrap with ordinary conformal blocks and / or no $\Delta_{T^\prime}$ gap is shown with a dotted line for comparison.}
\label{fig:super}
\end{figure}

It should come as no surprise that we have not seen any islands yet. If a point satisfies crossing symmetry, unitarity and the shadow relation for some spin-2 gap $\Delta_T$, it clearly continues to satisfy these criteria when $\Delta_T$ is made less restrictive. To produce an island in this situation, one must resort to imposing whichever additional gaps appear to be most plausible \cite{ls17b}. It is natural to ask if an LRI island can be produced by applying this logic in the spin-2 sector --- \textit{i.e.} by setting the continuum to begin at some $\Delta_{T^\prime} > \Delta_T$ so that the leading spin-2 operator is isolated.

It turns out that this problem demonstrates the power of the superblocks \eqref{superblocks}. Spectral plots analogous to Figure \ref{fig:scalar3} tell us that the region carved out by ordinary conformal blocks is perfectly compatible with a large spin-2 gap. It is only the extra OPE relations encoded by the superblocks that ensure a more restrictive region as $\Delta_{T^\prime}$ is increased. Figure \ref{fig:super} shows our attempt to isolate the $\Delta_T = 3.1$ LRI by imposing $\Delta_{T^\prime} \geq 4.5$. The result is a fairly large island in which the perturbative prediction from \cite{brrz17a} can be found near the bottom.\footnote{The boundary obtained by using superblocks without any $\Delta_{T^\prime}$ gap is essentially the same as what we would get from the ordinary blocks. This is no longer true for the higher values of $\Delta_T$ that lead to lobed regions in Figure \ref{fig:6corr} and Figure \ref{fig:zoom}.} There are most likely other allowed points outside this island that we have not attempted to find. It would be interesting to check how small the spin-2 gap can be made before the two regions reunite.

\section{Conclusion}
\label{sec:conclusion}
This work has been concerned with nonlocal 3D conformal field theories having thee relevant scalar primaries --- one $\mathbb{Z}_2$-even and two $\mathbb{Z}_2$-odd. We have computed numerical bootstrap constraints that follow from one, three and six correlators, finding progressively more interesting regions each time. These regions are distinguished by $\Delta_T$, the minimum allowed dimension of the leading spin-2 operator which appears to be a good proxy for the line of long-range Ising models. We have found strong evidence that the $\Delta_T \geq 3.5$ plot is saturated by the $s = \frac{d}{2}$ LRI at its leftmost edge. Evidence of the $\Delta_T \geq 3$ plots being saturated by $s = s_*$ has been clear since the work of \cite{kps14}. For intermediate values of $\Delta_T$, where all progress has been perturbative up until now, our six correlator results roughly separate into two regimes. The $\Delta_T > 3.25$ regions show a tendency for lobes to form with the left edge converging to the position of a long-range Ising model. To sharpen these lobes into kinks that fix the first few digits of $\left ( \Delta_\sigma, \Delta_\epsilon \right )$, it is probably enough to search for functionals in a very large search space using a combination of the methods in this paper. Unfortunately, cornering the $\Delta_T < 3.25$ models seems to be in a different category of difficulty.

Discussing the regime treated by Figure \ref{fig:zoom} first, some room for improvement is purely numerical. Excluding the largest possible area, given our spectral assumptions, has been prohibitive due to the sheer number of points that must be tested when we are interested in a fixed line and several values of $\Delta_T$. This prevents us from pushing our numerics to the number of derivative components considered in \cite{s15} for instance. This is unforunate considering that points with large external dimensions like $\left ( \Delta_\sigma, \Delta_\epsilon \right ) = \left ( \frac{3}{4}, \frac{3}{2} \right )$ are exactly where high derivative orders are needed most. Even Figure \ref{fig:1corr}, which goes up to $(m_{\mathrm{max}}, n_{\mathrm{max}}) = (7, 9)$, has some visible lack of convergence near this point. While this property of the numerical bootstrap is well known \cite{pptvv16}, it is worth pointing out that the situation can sometimes be reversed in analytic approaches \cite{m17, mp18}. The route towards constraining the $\Delta_T = 3.4$ LRI is close to the limit of what has been done but straightforward in principle. One should simply repeat the scan in Figure \ref{fig:zoom} with a very large number of derivatives and use the superblocks \eqref{superblocks} for every spin. The superblocks in particular have led to interesting bounds at $\Delta_T = 3.1$ where most of our other results are disappointing. This makes it tempting to apply them in the lobed regions as well. It could be useful to impose certain gaps beyond $\Delta_T \geq 3.4$ following \cite{ls17b}. However, our ability to do this at $\Delta_T = 3.4$ is limited since the spectrum of a generalized free theory is nowhere near as sparse as the spectrum of the SRI.

Our results for $\Delta_T < 3.25$ have not been sufficient to test the anomalous dimensions computed in \cite{brrz17a, brrz17b}. The one basic feature of this duality that we have confirmed in this work is the decoupling of $\chi$ at $s = s_*$ as seen in Figure \ref{fig:decoupling}. Without any highly suggestive regions to work with, we are left wondering whether the situation could be improved by a mixed correlator bootstrap that includes spinning operators as well. While this idea was proposed in \cite{dptv17, dkkps18}, the version applicable to the LRI would be much more difficult since none of its spinning operators are conserved. The same is true when considering the long-range $O(N)$ generalization. These models, which have more internal symmetry, also suffer from a lack of conserved currents. We can see this by noticing that the operator $\sigma^{[i} \chi^{j]}$ has the right quantum numbers to recombine with the would-be current $J^{[ij]}_\mu$ at the short-range end.

Viewing \cite{brrz17a} as a general recipie for constructing nonlocal CFTs, we expect that several theories other than the 3D LRI can be analyzed with the techniques developed here. We have already mentioned the 3D long-range $O(N)$ model defined by a straightforward generalization of \eqref{lri-hamiltonian}. Another interesting possibility would be exploring long-range fixed points having additional discrete symmetry along the lines of \cite{s18, ks18b}. A very concrete problem for the future would be returning to the 2D long-range Ising model which this work has only treated at the level of a single correlator. Our reason for avoiding a mixed correlator analysis here is that an exclusion plot with two relevant primaries can only become more permissive once a third such operator is added. As emphasized in \cite{b18}, the interesting part of the exclusion plot for 2D CFTs is identical in the one-correlator and three-correlator cases, even before we allow for the presence of a third relevant primary. Moreover, a large $\mathbb{Z}_2$-even gap, which is known to yield an island around the 2D (short-range) Ising model, is not applicable to the LRI due to the operator $\sigma\chi$. It thus appears that a successful bootstrap of the 2D LRI should include four-point functions with the non-conserved spin-2 operator $T_{\mu\nu}$. While this problem is more numerically intensive than anything considered in this work, it is clearly easier than the 3D version suggested above. The necessary blocks have been worked out and they in fact have a simple closed form \cite{o12}. It is possible that results from these blocks could resolve a discrepancy between the Monte Carlo simulations \cite{bpr13} and \cite{apr14}. These groups have both computed the critical exponent which we have called $\Delta_\epsilon$ at $s = 1.6$ and found results that do not agree within uncertainty. Whether $T_{\mu\nu}$ is used as an external operator in $d = 2$ or $d = 3$, we should point out that this raises the possibility of imposing only the minimal spectral assumption and still producing islands like Figure \ref{fig:super}.

The fundamental result which makes the LRI bootstrap possible is the nonlocal equation of motion. The most general analytic constraint that we have extracted from it so far is a quadratic OPE coefficient relation with four traceless symmetric primaries $\mathcal{O}_i$. It takes the form
\begin{equation}
\frac{\lambda^{(m)}_{12\chi} \lambda^{(n)}_{34\sigma}}{\lambda^{(m)}_{12\sigma} \lambda^{(n)}_{34\chi}} = \frac{R^{(m)}_{12}}{R^{(n)}_{34}} \label{general-ratio}
\end{equation}
with the right hand side given by \eqref{ratio-general}. This relation allows one to use techniques from the superconformal bootstrap, especially after setting $\mathcal{O}_1 = \sigma$, $\mathcal{O}_2 = \mathcal{O}$, $\mathcal{O}_3 = \chi$ and $\mathcal{O}_4 = \mathcal{O}$ (which together imply $m = n = 0$).\footnote{The version with general $(m, n)$ will be important for the larger correlator system that puts $T_{\mu\nu}$ on the same footing as $\sigma$, $\epsilon$ and $\chi$.} It was remarked in \cite{brrz17a} that nonlocality provides one with a degree of analytic control which is otherwise elusive in non-supersymmetric theories. In this work, we have performed a first analysis of the applicable superblocks \eqref{superblocks}. To strengthen the analogy between the LRI bootstrap and the superconformal bootstrap, one should seek a better understanding of the protected operators $[\sigma\chi]_{n, \ell}$.

First off, we are not aware of any perturbative results about odd-spin operators in the LRI other than $[\sigma\chi]_{0, 1}$. Even our anomalous dimension \eqref{dim-t} for the even-spin operator $T_{\mu\nu}$ appears to be new. Instead of evaluating increasingly complicated diagrams, we have arrived at the protected tower by noticing that continuity in $s$ requires odd-spin double-twist operators to stay at the pole \eqref{super-ratio-poles}. A more intuitive argument is that the bootstrap of a nonlocal CFT should not be significantly more powerful than the bootstrap of a local CFT. In the three-correlator system used for 3D CFTs since \cite{kps14}, all exchanged operators with odd spins are $\mathbb{Z}_2$-odd. Therefore, a bootstrap including all of the constraints related to $\epsilon$ and $\sigma$ (and hence $\partial^2 \sigma$) is not able to pick out a spectrum of odd-spin $\mathbb{Z}_2$-even operators at generic non-trivial dimensions.\footnote{While these operators have never been accessed by the numerical bootstrap, results about $\mathbb{Z}_2$-even vectors in Monte Carlo simulations recently appeared in \cite{mrly18}.} There is some justice to the fact that our bootstrap, based on the constraints of $\epsilon$, $\sigma$ and $\partial^s \sigma$, is not able to do so either. The connection between this tower and the superconformal bootstrap literature comes from the fact that we have only fixed the dimensions of the $[\sigma\chi]_{n, \ell}$ operators. If we could fix their squared OPE coefficients as well, we would be able to resum the full odd-spin contribution to $\left < \sigma\sigma\chi\chi \right >$ in what has become known as the minibootstrap \cite{brv15}.

We do not get access to these coefficients simply by recognizing that $\lambda_{\sigma\sigma\mathcal{O}} = \lambda_{\chi\chi\mathcal{O}} = 0$ when $\mathcal{O}$ has odd spin. Solving for $\lambda^2_{\sigma\chi\mathcal{O}}$, at least using the methods explored here, requires a more refined statement about how the Bose symmetric OPE coefficients vanish as the spin of $\mathcal{O}$ approaches an odd integer. Spin, in this context, can be defined as the homogeneity of an operator with respect to its polarization vector:
\begin{equation}
\mathbb{O}(X, \lambda Z) = \lambda^{\ell} \mathbb{O}(X, Z) \;\; , \;\; \ell \in \mathbb{C} \; . \label{continuous-spin}
\end{equation}
A detailed analysis of continuous spin trajectories for conformal field theories in Minkowski space recently appeared in \cite{ks18a}. The appendix of that paper computes the three-point tensor structures carried by the Wightman functions
\begin{equation}
\left < \mathcal{O}_1(X_1, Z_1) \mathbb{O}(X, Z) \mathcal{O}_2(X_2, Z_2) \right > \; , \; \left < \mathcal{O}_2(X_2, Z_2) \mathbb{O}(X, Z) \mathcal{O}_1(X_1, Z_1) \right > \label{wightman-functions}
\end{equation}
which involve two local operators along with $\mathbb{O}$ which is necessarily nonlocal. An important result is that the other possible orderings all vanish. This in particular means that correlators involving $\mathbb{O}$ depend on the operator ordering even at spacelike separation. It is therefore not possible to define Euclidean correlators by analytically continuing \eqref{wightman-functions} from the spacelike region \cite{ks18a}. As the numerical bootstrap operates on a Euclidean configuration, this lack of a clear prescription is an obstacle to implementing the minibootstrap for the crossing equation \eqref{app-copy}. Even if we had a continuous spin generalization of \eqref{super-ratio}, it would not necessarily be clear that $\lambda_{\sigma\sigma\mathcal{O}}$ and $\lambda_{\chi\chi\mathcal{O}}$ approach zero in a universal way. Perturbative expressions could therefore be helpful in sharpening our expectations.

The speculation that long-range Ising models could be bootstrapped first appeared in \cite{epprsv12}. The proper treatment of conformal invariance in \cite{prvz16} and the infrared duality in \cite{brrz17a, brrz17b} both lent subsequent support to this idea. In this work, we have aimed to study the space of nonlocal 3D CFTs in a way that imposes all non-perturbative results about the LRI that are currently known. This turns out to be a rich undertaking which has provided a new set of analytic relations in section \ref{sec:shadow} and interesting numerical regions in section \ref{sec:numerics}. The current state of our results can be summarized by a look at Figure \ref{fig:zoom} which shows candidate points for the LRI at various values of $s$ identified via the extremal functional method \cite{ps11, ep13, ep16}. While there is much room for optimizing this plot numerically, it already shows evidence that some long-range Ising models saturate the bounds from crossing symmetry and unitarity. The lobes where these models live can become quite narrow as $\Delta_T \rightarrow 3.5$ which describes the Gaussian fixed point. As we have discussed, there are opportunities for future progress that span several directions. Apart from more intensive numerics and perturbative checks covering more operators, it will be important to determine how the LRIs are restricted by analyticity in spin. Overall, bootstrap constraints on the space of nonlocal CFTs in $d > 1$ remain underexplored. This space is, in a sense, much larger than that of local CFTs. By fully exploiting the nonlocality of the LRI, we have shown that in one example, this dauntingly large space can still be narrowed down to the point of showing us interesting models.

\acknowledgments
Throughout this work I have benefited enormously from discussions with Leonardo Rastelli, Slava Rychkov and Bernardo Zan. I also thank Sheer El-Showk for collaboration during the early stages. Some helpful discussions with Damon Binder and Cathelijne ter Burg took place at the Bootstrap 2018 workshop in Caltech while this draft was in preparation. I am grateful to the organizers of Non-perturbative and Numerical Approaches to Quantum Gravity, String Theory and Holography in the ICTS, Bangalore for inviting me to present early results from this project. Travel to and from this workshop was funded by the Stony Brook Graduate Student Organization. This work was partially supported by the Natural Sciences and Engineering Research Council of Canada (CGS03-460190-2014). The numerical calculations were done on the SeaWulf cluster of the Institute for Advanced Computational Science which was made possible by National Science Foundation grant 1531492.

\appendix
\section{Implementation and conventions}
\label{sec:appa}
\subsection{Rational approximations}
To approximate conformal blocks for the numerical bootstrap, it is most efficient to use the radial co-ordinate $\rho$ \cite{hr13}. This is defined by using a conformal transformation to map our four points to the Euclidean configuration $\left < \phi_i(-1) \phi_j(-\rho, -\bar{\rho}) \phi_k(\rho, \bar{\rho}) \phi_l(1) \right >$. Writing $r = |\rho|$ and $\eta = \cos \arg \rho$, we convert the block to a meromorphic function of $\Delta$ and apply the recursion relation found in \cite{kps13}:
\begin{eqnarray}
H^{\Delta_{ij}, \Delta_{kl}}_{\Delta, \ell}(r, \eta) &=& r^{-\Delta} G^{\Delta_{ij}, \Delta_{kl}}_{\Delta, \ell}(r, \eta) \nonumber \\
H^{\Delta_{ij}, \Delta_{kl}}_{\Delta, \ell}(r, \eta) &=& H^{\Delta_{ij}, \Delta_{kl}}_{\infty, \ell}(r, \eta) + \sum_i \frac{c_i^{\Delta_{ij}, \Delta_{kl}} r^{n_i}}{\Delta - \Delta_i(\ell)} H^{\Delta_{ij}, \Delta_{kl}}_{\Delta_i(\ell) + n_i, \ell_i}(r, \eta) \; . \label{meromorphic-recursion}
\end{eqnarray}
\begin{table}[h]
\centering
\begin{tabular}{c|c|c|c}
$n_i$ & $\Delta_i(\ell)$ & $\ell_i$ & $c_i^{\Delta_{12}, \Delta_{34}}(\ell)$ \\
\hline
$k$ & $1 - \ell - k$ & $\ell + k$ & $c_1^{\Delta_{12}, \Delta_{34}}(\ell, k)$ \\
$2k$ & $1 + \nu - k$ & $\ell$ & $c_2^{\Delta_{12}, \Delta_{34}}(\ell, k)$ \\
$k$ & $1 + \ell + 2\nu - k$ & $\ell - k$ & $c_3^{\Delta_{12}, \Delta_{34}}(\ell, k)$
\end{tabular}
\caption{The data required to apply \eqref{meromorphic-recursion} where $\nu = \frac{d - 2}{2}$. Each series of poles is labelled by $k$ which in practice needs to be truncated at some $k_{\mathrm{max}}$. There is also a maximum spin in our problem which we label $\ell_{\mathrm{max}}$.}
\label{pole-data}
\end{table}
The entire piece and the residue expression, both found in \cite{kps14}, are:
\begin{eqnarray}
h_{\infty, \ell}^{\Delta_{12}, \Delta_{34}}(r, \eta) &=& \frac{\ell!}{(2\nu)_{\ell}} \frac{(-1)^{\ell} C_{\ell}^{\nu}(\eta) (1 - r^2)^{-\nu}}{(1 + r^2 + 2r\eta)^{\frac{1}{2}(1 + \Delta_{12} - \Delta_{34})}(1 + r^2 - 2r\eta)^{\frac{1}{2}(1 - \Delta_{12} + \Delta_{34})}} \label{longer-data} \\
c_1^{\Delta_{12}, \Delta_{34}}(\ell, k) &=& -\frac{k(-4)^k}{(k!)^2} \frac{(\ell + 2\nu)_k}{(\ell + \nu)_k} \left ( \frac{1}{2} (1 - k + \Delta_{12}) \right )_k \left ( \frac{1}{2} (1 - k + \Delta_{34}) \right )_k \nonumber \\
c_2^{\Delta_{12}, \Delta_{34}}(\ell, k) &=& \frac{k (\nu + 1)_{k - 1} (-\nu)_{k + 1}}{(k!)^2} \frac{\ell + \nu - k}{\ell + \nu + k} \left ( \frac{\ell + \nu - k + 1}{2} \right )_k^{-2} \left ( \frac{\ell + \nu - k}{2} \right )_k^{-2} \nonumber \\
&& \left ( \frac{1}{2} (1 - k + \ell - \Delta_{12} + \nu) \right )_k \left ( \frac{1}{2} (1 - k + \ell + \Delta_{12} + \nu) \right )_k \nonumber \\
&& \left ( \frac{1}{2} (1 - k + \ell - \Delta_{34} + \nu) \right )_k \left ( \frac{1}{2} (1 - k + \ell + \Delta_{34} + \nu) \right )_k \nonumber \\
c_3^{\Delta_{12}, \Delta_{34}}(\ell, k) &=& -\frac{k(-4)^k}{(k!)^2} \frac{(\ell + 1 - k)_k}{(\ell + \nu + 1 - k)_k} \left ( \frac{1}{2} (1 - k + \Delta_{12}) \right )_k \left ( \frac{1}{2} (1 - k + \Delta_{34}) \right )_k \; . \nonumber
\end{eqnarray}
With this data, \eqref{meromorphic-recursion} can be used to generate derivatives with respect to $(z, \bar{z})$ defined such that $u = |z|^2$ and $v = |1 - z|^2$. However, a more efficient approach is to switch to the variables $a = z + \bar{z}$ and $b = (z - \bar{z})^2$ and compute derivatives around $a = 1$. Derivatives around $b = 0$ can then be found from the second-order Casimir differential equation satisfied by the blocks as explained in \cite{hor13}. Due to the structure of this equation, the pattern exhibited by $\partial_a^m \partial_b^n G^{\Delta_{ij}, \Delta_{kl}}_{\Delta, \ell}(a, b)$ is:
\begin{eqnarray}
n &\in& \{ 0, \dots, n_{\mathrm{max}} \} \nonumber \\
m &\in& \{ 0, \dots, 2(n_{\mathrm{max}} - n) + m_{\mathrm{max}} \} \; . \label{ab-derivs}
\end{eqnarray}
Our choices for $(m_{\mathrm{max}}, n_{\mathrm{max}})$ are stated alongside the results in section \ref{sec:numerics}. The other cutoff parameters used in this work are $k_{\mathrm{max}} = 40$ and $\ell_{\mathrm{max}} = 20$.

If we wish to combine conformal blocks according to \eqref{superblocks}, the correct rational approximations will include extra poles apart from those captured in Table \ref{pole-data}. The coefficient $R(\Delta, \ell)$ in the superblock
\begin{equation}
\mathcal{G}_{\Delta, \ell}(u, v) = G_{\Delta, \ell}^{0, 0}(u, v) + R(\Delta, \ell) v^{\Delta_\sigma - \frac{d}{2}} G_{\Delta, \ell}^{\Delta_{\chi\sigma}, \Delta_{\sigma\chi}}(u, v) \label{repeated-block}
\end{equation}
can be written as an infinite product since the numerator and denominator of \eqref{super-ratio-expression} have the same sum of gamma function arguments.
\begin{equation}
R(\Delta, \ell) = \prod_{k = 0}^{\infty} \frac{(\Delta + \ell + 2k)^2 (\Delta - 2\Delta_\sigma - \ell - 2k) (\Delta + 2\Delta_\sigma - 2d - \ell - 2k)}{(\Delta - d - \ell - 2k)^2 (\Delta - 2\Delta_\sigma + d + \ell + 2k) (\Delta + 2\Delta_\sigma - d + \ell + 2k)} \label{fortunate-product}
\end{equation}

To rule out solutions to crossing, we use the semidefinite program solver \texttt{SDPB} \cite{s15}. This requires a ``positive-times-polynomial'' expression for the derivative of each conformal block, or more precisely, convolved conformal block \eqref{convolved-block} as these are what enter in crossing equations. To find these expressions with the above algorithm, we use the helper program \texttt{PyCFTBoot} \cite{b17}. Note that the existence of a positive prefactor for each polynomial is usually attributed to the fact that poles in Table \ref{pole-data} are below the unitarity bound. For the superblocks in this work, we have poles on the whole real line coming from \eqref{fortunate-product}. This time, positivity is a consequence of the fact that every pole above the unitarity bound is a double pole. Although this seems forunate, numerical stability is an additional obstacle to using the rational approximation \eqref{fortunate-product}. We will return to this issue at the end of this appendix.

\subsection{Crossing equations}
Here, we write the statement of crossing symmetry used to find kinks in this work. Clearly, $\left < \sigma\sigma\sigma\sigma \right >$, $\left < \epsilon\epsilon\epsilon\epsilon \right >$ and $\left < \chi\chi\chi\chi \right >$ each have one crossing equation. By repeating the analysis of \cite{kps14}, one sees that the mixed correlators $\left < \sigma\sigma\epsilon\epsilon \right >$, $\left < \sigma\sigma\chi\chi \right >$ and $\left < \epsilon\epsilon\chi\chi \right >$ each have three. The full system is
\begin{equation}
\sum_{\mathcal{O}^+_{2 | \ell}} \left ( \lambda_{\sigma\sigma\mathcal{O}} \; \lambda_{\epsilon\epsilon\mathcal{O}} \; \lambda_{\chi\chi\mathcal{O}} \right ) V^{(0)}_{\Delta, \ell} \left ( \begin{tabular}{c} $\lambda_{\sigma\sigma\mathcal{O}}$ \\ $\lambda_{\epsilon\epsilon\mathcal{O}}$ \\ $\lambda_{\chi\chi\mathcal{O}}$ \end{tabular} \right ) + \sum_{\mathcal{O}^-} \lambda^2_{\sigma\epsilon\mathcal{O}} V^{(1)}_{\Delta, \ell} + \lambda^2_{\epsilon\chi\mathcal{O}} V^{(3)}_{\Delta, \ell} + \sum_{\mathcal{O}^+} \lambda^2_{\sigma\chi\mathcal{O}} V^{(2)}_{\Delta, \ell} = 0 \label{6corr-crossing1}
\end{equation}
where each vector has twelve components. We first write these components for the sums with no restriction on spin.
\begin{equation}
V^{(1)}_{\Delta, \ell} = \left [ \begin{tabular}{c} $0$ \\ $0$ \\ $0$ \\ $F_{-,\Delta,\ell}^{\sigma\epsilon ; \sigma\epsilon}$ \\ $0$ \\ $0$ \\ $(-1)^\ell F_{-,\Delta,\ell}^{\epsilon\sigma ; \sigma\epsilon}$ \\ $-(-1)^\ell F_{+,\Delta,\ell}^{\epsilon\sigma ; \sigma\epsilon}$ \\ $0$ \\ $0$ \\ $0$ \\ $0$ \end{tabular} \right ],
V^{(2)}_{\Delta, \ell} = \left [ \begin{tabular}{c} $0$ \\ $0$ \\ $0$ \\ $0$ \\ $F_{-,\Delta,\ell}^{\sigma\chi ; \sigma\chi}$ \\ $0$ \\ $0$ \\ $0$ \\ $(-1)^\ell F_{-,\Delta,\ell}^{\chi\sigma ; \sigma\chi}$ \\ $-(-1)^\ell F_{+,\Delta,\ell}^{\chi\sigma ; \sigma\chi}$ \\ $0$ \\ $0$ \end{tabular} \right ],
V^{(3)}_{\Delta, \ell} = \left [ \begin{tabular}{c} $0$ \\ $0$ \\ $0$ \\ $0$ \\ $0$ \\ $F_{-,\Delta,\ell}^{\epsilon\chi ; \epsilon\chi}$ \\ $0$ \\ $0$ \\ $0$ \\ $0$ \\ $(-1)^\ell F_{-,\Delta,\ell}^{\chi\epsilon ; \epsilon\chi}$ \\ $-(-1)^\ell F_{+,\Delta,\ell}^{\chi\epsilon ; \epsilon\chi}$ \end{tabular} \right ] \label{6corr-crossing2}
\end{equation}
The first sum, whose operators must have even spin, involves components that are $3 \times 3$ matrices. We write them individually as
\begin{eqnarray}
&& V^{(0)}_1 = \left ( \begin{tabular}{ccc} $F_{-,\Delta,\ell}^{\sigma\sigma ; \sigma\sigma}$ & $0$ & $0$ \\ $0$ & $0$ & $0$ \\ $0$ & $0$ & $0$ \end{tabular} \right ) \; , \; V^{(0)}_{7, 8} = \left ( \begin{tabular}{ccc} $0$ & $\frac{1}{2} F_{\mp,\Delta,\ell}^{\sigma\sigma ; \epsilon\epsilon}$ & $0$ \\ $\frac{1}{2} F_{\mp,\Delta,\ell}^{\sigma\sigma ; \epsilon\epsilon}$ & $0$ & $0$ \\ $0$ & $0$ & $0$ \end{tabular} \right ) \nonumber \\
&& V^{(0)}_2 = \left ( \begin{tabular}{ccc} $0$ & $0$ & $0$ \\ $0$ & $F_{-,\Delta,\ell}^{\epsilon\epsilon ; \epsilon\epsilon}$ & $0$ \\ $0$ & $0$ & $0$ \end{tabular} \right ) \; , \; V^{(0)}_{9, 10} = \left ( \begin{tabular}{ccc} $0$ & $0$ & $\frac{1}{2} F_{\mp,\Delta,\ell}^{\sigma\sigma ; \chi\chi}$ \\ $0$ & $0$ & $0$ \\ $\frac{1}{2} F_{\mp,\Delta,\ell}^{\sigma\sigma ; \chi\chi}$ & $0$ & $0$ \end{tabular} \right ) \nonumber \\
&& V^{(0)}_3 = \left ( \begin{tabular}{ccc} $0$ & $0$ & $0$ \\ $0$ & $0$ & $0$ \\ $0$ & $0$ & $F_{-,\Delta,\ell}^{\chi\chi ; \chi\chi}$ \end{tabular} \right ) \; , \; V^{(0)}_{11, 12} = \left ( \begin{tabular}{ccc} $0$ & $0$ & $0$ \\ $0$ & $0$ & $\frac{1}{2} F_{\mp,\Delta,\ell}^{\epsilon\epsilon ; \chi\chi}$ \\ $0$ & $\frac{1}{2} F_{\mp,\Delta,\ell}^{\epsilon\epsilon ; \chi\chi}$ & $0$ \end{tabular} \right ) \label{6corr-crossing3}
\end{eqnarray}
with the rest being zero.

As we have already mentioned, these crossing equations do not account for the correlators $\left < \sigma\sigma\sigma\chi \right >$, $\left < \chi\chi\chi\sigma \right >$ and $\left < \epsilon\epsilon\sigma\chi \right >$. Even though \eqref{6corr-crossing1} is enough to impose both constraints from the nonlocal equation of motion, the more general system makes it clear that we do not need multiple sums for operators in the same representation. The sum rule
\begin{eqnarray}
&& \sum_{\mathcal{O}^+_{2 | \ell}} \left ( \lambda_{\sigma\sigma\mathcal{O}} \; \lambda_{\epsilon\epsilon\mathcal{O}} \; \lambda_{\chi\chi\mathcal{O}} \; \lambda_{\sigma\chi\mathcal{O}} \right ) \tilde{V}^{(0)}_{\Delta, \ell} \left ( \begin{tabular}{c} $\lambda_{\sigma\sigma\mathcal{O}}$ \\ $\lambda_{\epsilon\epsilon\mathcal{O}}$ \\ $\lambda_{\chi\chi\mathcal{O}}$ \\ $\lambda_{\sigma\chi\mathcal{O}}$ \end{tabular} \right ) \nonumber \\
&& + \sum_{\mathcal{O}^-} \left ( \lambda_{\sigma\epsilon\mathcal{O}} \; \lambda_{\epsilon\chi\mathcal{O}} \right ) \tilde{V}^{(1)}_{\Delta, \ell} \left ( \begin{tabular}{c} $\lambda_{\sigma\epsilon\mathcal{O}}$ \\ $\lambda_{\epsilon\chi\mathcal{O}}$ \end{tabular} \right ) + \sum_{\mathcal{O}^+_{2 \nmid \ell}} \lambda^2_{\sigma\chi\mathcal{O}} \tilde{V}^{(2)}_{\Delta, \ell} = 0 \label{9corr-crossing1}
\end{eqnarray}
now has sixteen rows. It is trivial to determine the first twelve by demanding that the equations of \eqref{6corr-crossing1} are captured in \eqref{9corr-crossing1}. Therefore, we will simply write
\begin{eqnarray}
&& \tilde{V}^{(0)}_{13} = \left ( \begin{tabular}{cccc} $0$ & $0$ & $0$ & $\frac{1}{2} F_{-,\Delta,\ell}^{\sigma\sigma ; \sigma\chi}$ \\ $0$ & $0$ & $0$ & $0$ \\ $0$ & $0$ & $0$ & $0$ \\ $\frac{1}{2} F_{-,\Delta,\ell}^{\sigma\sigma ; \sigma\chi}$ & $0$ & $0$ & $0$ \end{tabular} \right ) , \tilde{V}^{(0)}_{14} = \left ( \begin{tabular}{cccc} $0$ & $0$ & $0$ & $0$ \\ $0$ & $0$ & $0$ & $0$ \\ $0$ & $0$ & $0$ & $\frac{1}{2} F_{-,\Delta,\ell}^{\chi\chi ; \sigma\chi}$ \\ $0$ & $0$ & $\frac{1}{2} F_{-,\Delta,\ell}^{\chi\chi ; \sigma\chi}$ & $0$ \end{tabular} \right ) \nonumber \\
&& \tilde{V}^{(0)}_{15} = \left ( \begin{tabular}{cccc} $0$ & $0$ & $0$ & $0$ \\ $0$ & $0$ & $0$ & $\frac{1}{2} F_{-,\Delta,\ell}^{\epsilon\epsilon ; \sigma\chi}$ \\ $0$ & $0$ & $0$ & $0$ \\ $0$ & $\frac{1}{2} F_{-,\Delta,\ell}^{\epsilon\epsilon ; \sigma\chi}$ & $0$ & $0$ \end{tabular} \right ) , \tilde{V}^{(0)}_{16} = \left ( \begin{tabular}{cccc} $0$ & $0$ & $0$ & $0$ \\ $0$ & $0$ & $0$ & $\frac{1}{2} F_{+,\Delta,\ell}^{\epsilon\epsilon ; \sigma\chi}$ \\ $0$ & $0$ & $0$ & $0$ \\ $0$ & $\frac{1}{2} F_{+,\Delta,\ell}^{\epsilon\epsilon ; \sigma\chi}$ & $0$ & $0$ \end{tabular} \right ) \label{9corr-crossing2}
\end{eqnarray}
for $\mathcal{O}^+_{2 | \ell}$,
\begin{equation}
\tilde{V}^{(1)}_{13, 14} = \left ( \begin{tabular}{cc} $0$ & $0$ \\ $0$ & $0$ \end{tabular} \right ) \; , \; \tilde{V}^{(1)}_{15, 16} = \left ( \begin{tabular}{cc} $0$ & $\frac{1}{2} F_{\mp,\Delta,\ell}^{\sigma\epsilon ; \epsilon\chi}$ \\ $\frac{1}{2} F_{\mp,\Delta,\ell}^{\sigma\epsilon ; \epsilon\chi}$ & $0$ \end{tabular} \right ) \label{9corr-crossing3}
\end{equation}
for $\mathcal{O}^-$ and $\tilde{V}^{(2)}_{13, 14, 15, 16} = 0$ for $\mathcal{O}^+_{2 \nmid \ell}$.
The presence of extra equations and larger matrices certainly increases the computation time when using the full system \eqref{9corr-crossing1}. However, the main impact on performance comes from the fact that conformal blocks with vanishing and non-vanishing dimension differences can now be found in the same matrix. In simpler problems, the former blocks can be approximated with rational functions of a much lower degree, owing to the fact that their residues vanish for half of the poles in the Table \ref{pole-data}. In the present case, this advantage is lost as all entries of the $4 \times 4$ matrix need to be accompanied by a common denominator.

After trying a few examples, the bounds from \eqref{9corr-crossing1} appear indistinguishable from those obtained with \eqref{6corr-crossing1}. This is consistent with a piece of lore stating that additional four-point functions only help if they increase the number of gaps / OPE constraints that can be imposed. An earlier example of this was noticed in \cite{kps14} which first studied the 3D Ising model using three correlators. Without a constraint on the number of relevant $\mathbb{Z}_2$-odd operators, the allowed region turned out to be the same as what was already known from a single correlator.

\subsection{Discretization for superblocks}
This common denominator mentioned in the ten correlator bootstrap appears in the six correlator case as well if the equations are formulated with superblocks. It also needs to be extended to include the poles of \eqref{fortunate-product}, leading to a four-fold increase in the degree of each rational approximation. An even more severe problem concerns the stability of \texttt{SDPB}. For normal operation, the user constructs a measure out of a given block's poles and then uses it to compute a basis of orthogonal polynomials \cite{s15}. These polynomials only exist for the ordinary blocks because our superblocks are singular above the unitarity bound. In other words, the double poles do not interfere with positivity but they lead to a measure that is not normalizable.

Because of this, we have been unable to complete a full run of \texttt{SDPB} by applying $k_{\mathrm{max}} = 40$ to the rational approximation \eqref{fortunate-product}. Instead, Figure \ref{fig:super} was produced by evaluating superblocks on a discrete $\Delta$ grid which allows us to use the exact gamma functions. It would be interesting to see if alternative solvers are able to remove this complication. Table \ref{discretization} shows our discretization choices for odd-spin operators (which are protected) and even-spin operators (which should approximate a continuum) when using the information in \eqref{change2}.

\begin{table}[h]
\centering
\begin{tabular}{c|c|c}
& Step size & Number of points \\
\hline
$\ell = 0$ & $0.01$ & $1700$ \\
$\ell = 2$ & $0.01$ & $1550$ \\
$2 \nmid \ell$ & $2$ & $10$
\end{tabular}
\caption{The grid spacing $\Delta_{N + 1} - \Delta_N$ and number of points used for operators in $\sigma \times \chi$ once all consequences of the shadow relation are imposed. After the last point, we use rational approximation to demand positivity on individual sum rule vectors in a continuum. In other words, we no longer impose the shadow relation after a high enough cutoff in $\Delta$.}
\label{discretization}
\end{table}
This results in file sizes of about 1GB. We have checked that shrinking the step size and including superblocks for $\ell = 4$ does not lead to a significant growth in the allowed region of Figure \ref{fig:super}.

\bibliographystyle{utphys}
\bibliography{references}
\end{document}